\documentclass[twocolumn,tighten,times,]{aastex63}
\usepackage{graphicx}
\usepackage{subfigure}
\usepackage{amsmath}
\usepackage{amssymb}
\usepackage{natbib}
\usepackage{sidecap}
\usepackage{cancel}
\usepackage{multirow}
\usepackage{stmaryrd}
\usepackage{float}
\usepackage[T1]{fontenc}
\def\jref@jnl#1{{\rm#1}}

\def\actaa{\jref@jnl{Acta Astron.}}      
\def\aj{\jref@jnl{AJ}}                   
\def\araa{\jref@jnl{ARA\&A}}             
\def\apj{\jref@jnl{ApJ}}                 
\def\apjl{\jref@jnl{ApJ}}                
\def\apjs{\jref@jnl{ApJS}}               
\def\ao{\jref@jnl{Appl.~Opt.}}           
\def\apss{\jref@jnl{Ap\&SS}}             
\def\aap{\jref@jnl{A\&A}}                
\def\aapr{\jref@jnl{A\&A~Rev.}}          
\def\aaps{\jref@jnl{A\&AS}}              
\def\azh{\jref@jnl{AZh}}                 
\def\baas{\jref@jnl{BAAS}}               
\def\jrasc{\jref@jnl{JRASC}}             
\def\jcap{\jref@jnl{JCAP}}               
\def\memras{\jref@jnl{MmRAS}}            
\def\mnras{\jref@jnl{MNRAS}}             
\def\na{\jref@jnl{New Astronomy}}        
\def\pra{\jref@jnl{Phys.~Rev.~A}}        
\def\prb{\jref@jnl{Phys.~Rev.~B}}        
\def\prc{\jref@jnl{Phys.~Rev.~C}}        
\def\prd{\jref@jnl{Phys.~Rev.~D}}        
\def\pre{\jref@jnl{Phys.~Rev.~E}}        
\def\prl{\jref@jnl{Phys.~Rev.~Lett.}}    
\def\pasa{\jref@jnl{PASA}}               
\def\pasp{\jref@jnl{PASP}}               
\def\pasj{\jref@jnl{PASJ}}               
\def\qjras{\jref@jnl{QJRAS}}             
\def\rmxaa{\jref@jnl{Rev.Mex.AA}}       
\def\skytel{\jref@jnl{S\&T}}             
\def\solphys{\jref@jnl{Sol.~Phys.}}      
\def\sovast{\jref@jnl{Soviet~Ast.}}      
\def\ssr{\jref@jnl{Space~Sci.~Rev.}}     
\def\zap{\jref@jnl{ZAp}}                 
\def\nar{\jref@jnl{NewAR}}               
\def\nat{\jref@jnl{Nature}}              
\def\iaucirc{\jref@jnl{IAU~Circ.}}       
\def\aplett{\jref@jnl{Astrophys.~Lett.}} 
\def\apspr{\jref@jnl{Astrophys.~Space~Phys.~Res.}}
\def\bain{\jref@jnl{Bull.~Astron.~Inst.~Netherlands}} 
\def\fcp{\jref@jnl{Fund.~Cosmic~Phys.}}  
\def\gca{\jref@jnl{Geochim.~Cosmochim.~Acta}}   
\def\grl{\jref@jnl{Geophys.~Res.~Lett.}} 
\def\jcap{\jref@jnl{JCAP}}      %
\def\jcp{\jref@jnl{J.~Chem.~Phys.}}      
\def\jgr{\jref@jnl{J.~Geophys.~Res.}}    
\def\jqsrt{\jref@jnl{J.~Quant.~Spec.~Radiat.~Transf.}}
\def\memsai{\jref@jnl{Mem.~Soc.~Astron.~Italiana}}
\def\nphysa{\jref@jnl{Nucl.~Phys.~A}}   
\def\physrep{\jref@jnl{Phys.~Rep.}}   
\def\physscr{\jref@jnl{Phys.~Scr}}   
\def\planss{\jref@jnl{Planet.~Space~Sci.}}   
\def\procspie{\jref@jnl{Proc.~SPIE}}   

\defcitealias{Sironi_20}{SB20}
\defcitealias{Sridhar+21c}{SSB21}
\defcitealias{Sridhar+23a}{SSB23}

\newcommand{\eqb}{\begin{eqnarray}}
\newcommand{\eqe}{\end{eqnarray}}
\newcommand{\be}{\begin{eqnarray}}
\newcommand{\ee}{\end{eqnarray}}
\newcommand{\bi}{\begin{itemize}}
\newcommand{\ei}{\end{itemize}}

\begin{document}

\title{Bulk Motions in the Black Hole Jet Sheath as a Candidate for the Comptonizing Corona}

\author[0000-0002-5519-9550]{Navin Sridhar}
\affiliation{Department of Astronomy, Columbia University, New York, NY 10027, USA}
\affiliation{Theoretical High Energy Astrophysics (THEA) Group, Columbia University, New York, NY 10027, USA}
\affiliation{Cahill Center for Astronomy and Astrophysics, California Institute of Technology, Pasadena, CA 91106, USA}
\affiliation{Department of Physics, Stanford University, 382 Via Pueblo Mall, Stanford, CA 94305, USA}
\affiliation{Kavli Institute for Particle Astrophysics \& Cosmology, P.O. Box 2450, Stanford University, Stanford, CA 94305, USA}
\correspondingauthor{Navin Sridhar}
\email{nsridhar@stanford.edu}

\author[0000-0002-7301-3908]{Bart Ripperda}
\affiliation{Canadian Institute for Theoretical Astrophysics, 60 St. George St, Toronto, ON M5S 3H8, Canada}
\affiliation{Department of Physics, University of Toronto, 60 St. George St, Toronto, ON M5S 1A7, Canada}
\affiliation{David A. Dunlap Department of Astronomy \& Astrophysics, University of Toronto, 50 St. George St, Toronto, ON M5S 3H4, Canada}
\affiliation{Perimeter Institute for Theoretical Physics, 31 Caroline St. North, Waterloo, ON N2L 2Y5, Canada}

\author[0000-0002-1227-2754]{Lorenzo Sironi}
\affiliation{Department of Astronomy, Columbia University, New York, NY 10027, USA}
\affiliation{Theoretical High Energy Astrophysics (THEA) Group, Columbia University, New York, NY 10027, USA}
\affiliation{Center for Computational Astrophysics, Flatiron Institute, 162 5th Ave, New York, NY 10010, USA}

\author[0000-0002-2685-2434]{Jordy Davelaar}
\altaffiliation{NASA Hubble Fellowship Program, Einstein Fellow}
\affiliation{Department of Astrophysical Sciences, Peyton Hall, Princeton University, Princeton, NJ 08544, USA}

\author[0000-0001-5660-3175]{Andrei M. Beloborodov}
\affiliation{Theoretical High Energy Astrophysics (THEA) Group, Columbia University, New York, NY 10027, USA}
\affiliation{Department of Physics, Columbia University, New York, NY 10027, USA}

\begin{abstract}
Using two-dimensional general relativistic resistive magnetohydrodynamic simulations, we investigate the properties of the sheath separating the black hole jet from the surrounding medium. We find that the electromagnetic power flowing through the jet sheath is comparable to the overall accretion power of the black hole. The sheath is an important site of energy dissipation as revealed by the copious appearance of reconnection layers and plasmoid chains. About 20\% of the sheath power is dissipated between 2 and 10 gravitational radii. The plasma in the dissipative sheath moves along a nearly paraboloidal surface with trans-relativistic bulk motions dominated by the radial component, whose dimensionless 4-velocity is $\sim1.2\pm0.5$. In the frame moving with the mean (radially-dependent) velocity, the distribution of stochastic bulk motions resembles a Maxwellian with an `effective bulk temperature' of $\sim$100\,keV. Scaling the global simulation to Cygnus X-1 parameters gives a rough estimate of the Thomson optical depth across the jet sheath $\sim 0.01-0.1$, and it may increase in future magnetohydrodynamic simulations with self-consistent radiative losses. These properties suggest that the dissipative jet sheath may be a viable `coronal' region, capable of upscattering seed soft photons into a hard, nonthermal tail, as seen during the hard states of X-ray binaries and active galactic nuclei.
\end{abstract}


\section{Introduction}
\subsection{Microphysics} \label{subsec:microphysics}
The X-ray spectra of accreting black holes can feature a hard nonthermal X-ray tail with a cutoff at $\sim100$\,keV, predominantly seen during the `hard-states' of X-ray binaries \citep{1972ApJ...177L...5T, McClintock&Remillard_06, Done+07}. The traditional paradigm involves \textit{thermal Comptonization} of soft accretion disk photons by a bath of hot Maxwellian electrons in a `corona' with a temperature of $k_{\rm B}T_{\rm e}\sim100$\,keV \citep[][]{1976SvAL....2..191B, 1979Natur.279..506S, Zdziarski&Gierlinski_04}; the heating of the electrons, in this paradigm, is usually attributed to processes such as magnetic reconnection \citep{Galeev+79, Dimatteo, Merloni2001a, Merloni2001b} and turbulence \citep{Chandran+18}.  

Magnetic reconnection is incapable of sustaining a high electron temperature in the radiative corona \citep{belo_17}. It works in such a way that most of the plasma remains cool; however, the reconnection layer still efficiently Comptonizes photons by random bulk motions of the self-similar plasmoid chain created in the layer. This mechanism was found capable of producing the hard-state spectrum \citep{belo_17} and then demonstrated by first-principle, radiative, kinetic plasma (particle-in-cell; PIC) simulations of magnetic reconnection (\citealt{Sironi_20, Sridhar+21c, Sridhar+23a}, hereafter, \citetalias{Sironi_20}, \citetalias{Sridhar+21c}, \citetalias{Sridhar+23a}, see also \citealt{werner_18, Mehlhaff_20}).
The bulk motions of plasmoids/magnetic islands are generated as a natural outcome of magnetic reconnection and can achieve trans-relativistic speeds (with an energy distribution resembling a Maxwellian having an `effective bulk temperature' of $\sim100$\,keV). The stochastic bulk motions of this chain of plasmoids can Compton upscatter the soft X-ray photons to a hard, nonthermal tail. This was termed the \textit{cold-chain Comptonization} process, as most of the electrons that reside inside the chain of Comptonizing plasmoids are cooled down to non-relativistic temperatures due to inverse Compton (IC) losses. A similar process can also operate in turbulent plasmas where the bulk motions of the eddies can Compton upscatter soft photons to a nonthermal distribution (\citealt{Groselj+24, Nattila_2024}). 

\citetalias{Sironi_20}, \citetalias{Sridhar+21c}, and \citetalias{Sridhar+23a} investigated the dependence of the cold-chain Comptonization process on the IC cooling strength, the composition of the plasma (electron-positron vs. electron-ion) and the magnetization $\sigma$ of the reconnecting plasma ($\sigma\equiv B_0^2/(4\pi\rho c^2)$, where $B_0$ is the `upstream' magnetic field strength, $c$ the speed of light, and $\rho$ is the plasma density). The dependence of this process on the `guide field' ($B_{\rm g}$) i.e., the magnetic field that is oriented along the electric current in the reconnection layer, was investigated by \cite{Gupta+24}. These studies showed that the cold-chain Comptonization process can produce the observed nonthermal X-ray spectra for $\sigma \gtrsim 3$ and $B_{\rm g}/B_{\rm 0}\lesssim 0.3$, regardless of the composition of the corona. These were local box simulations that were performed with hierarchy-preserving scaled-down parameters (see \S 3 of \citetalias{Sridhar+23a} for more details). Furthermore, these simulations were performed assuming that the upstream plasma conditions are uniform across the current sheet (i.e., across the oppositely oriented magnetic field lines that reconnect), and being agnostic to the location and orientation of the reconnection layer with respect to the black hole and the accretion disk. The geometry and the location of the corona have remained elusive topics, which we aim to address in this work.

\subsection{Macrophysics}
X-ray spectral and temporal studies of black hole X-ray binaries have suggested that the location, size, and geometry of the corona could change depending on the phase of X-ray binary outburst \citep{Kara_19, Sridhar+19, Sridhar+20, Wang+21, Cao+21, Mendez+22}. Furthermore, the nonthermal X-ray emission during the hard states was typically seen contemporaneously with the presence of a compact ($\lesssim10^{15}\,{\rm cm}$) radio jet \citep{Fender&Kuulkers_01, Gallo+03, Fender+04b, Gallo+18, Mendez+22}. These observations have motivated models of coronae---albeit involving the less feasible thermal Comptonization process by hot electrons---that involve the jet or its base \citep{Markoff_05, Qiao+15, Lucchini+22}. Recent results from the Imaging X-ray Polarimetry Explorer may also help constrain the coronal geometry \citep[\textit{IXPE};][]{Weisskopf+16, Krawczynski+22, Marinucci+22, Ursini+22, Tagliacozzo+23}. 

The corona was proposed to outflow at a mildly relativistic speed from the accretion disk in order to explain the hard spectral index and the weak reflection in hard-state black holes \citep{Beloborodov_99, Malzac+01}. In addition, \citet{Beloborodov_98} showed that such outflows should generate polarization parallel to the disk normal. The picture of an outflowing corona has been used recently to explain the parallel polarization degree (PD) of $4.0\%\pm0.2\%$ observed by \textit{IXPE}  in the 2--8\,keV band from Cygnus X-1 \citep{Poutanen+23}. \cite{Dexter&Begelman24} further argued that the Comptonizing region might be a trans-relativistic outflow situated along the jet sheath. A similar picture has been proposed based on global radiative general relativistic magnetohydrodynamic (GRMHD) simulations \citep{Moscibrodzka_23}.

Global GRMHD simulations showing potential regions of magnetic dissipation can provide insights into the geometry and orientation of a putative corona, as well as the spatial and temporal variations in the polarization properties of the Comptonized emission \citep{Schnittman&Krolik_10, Beheshtipour+17}. Within the tenuous corona, magnetic dissipation could occur in a collisionless regime \citep{Goodman&Uzdensky_08}. However, performing global PIC simulations of a realistic accreting system that would accurately capture the global dynamics and kinetic dissipation scales is computationally unfeasible (although two-dimensional general relativistic PIC simulations by \citealt{Galishnikova+23} with simplifying assumptions on the accretion model, show that magnetic flux eruptions are a generic result of saturation of flux on the horizon). A simpler approach is based on GRMHD simulations with
explicit resistivity (e.g., \citealt{Qian+18, Vourellis+19, Ripperda+20}).\footnote{Although note that the magnetic reconnection rate in collisional plasma as described by MHD is about a tenth of the collisionless rate as found in PIC simulations} Such simulations allow modeling the global dynamics of accretion disks while resolving reconnection in the largest current sheets with a converged dissipation rate. GRMHD simulations have revealed the formation of current sheets and plasmoid-unstable reconnection layers in two regions: (1) the sheath at the interface between the accretion disk and the Poynting-flux-dominated polar jet \citep{Ripperda+20, Nathanail+20, Chashkina+21, Dihingia+22, Nathanail+22}, and (2) the magnetospheric region between the black hole and the inner accretion disk \citep{Ripperda+20, Ripperda+22} (see also a number of GRMHD and PIC simulations of the magnetospheric region without an accretion disk: \citealt{Parfrey+19, Inda-Koide+19, Bransgrove+21, Hirotani+21, Crinquand+22, El_Mellah+23}). 

In this paper, we investigate the structure of the dissipation region in the jet sheath using the high-resolution two-dimensional non-radiative general relativistic resistive magnetohydrodynamic (GRRMHD) simulation of \cite{Ripperda+20}.
The simulation was performed for the case of a magnetically arrested disk \citep[MAD;][]{Bisnovatyi-Kogan&Ruzmaikin_74, Bisnovatyi-Kogan&Ruzmaikin_76, Igumenshchev+03, Narayan+03}. MAD flows exhibit long phases of accretion where magnetic flux accumulates on the horizon and shorter eruptive phases where magnetic flux is ejected. Spinning black holes in the MAD state show the formation of strong jets, due to the large accumulation of flux. We examine the jet sheath during the MAD state as a candidate for the source of hard X-rays, determine the power available for dissipation, and investigate the statistics of plasma motions. These motions can potentially Comptonize seed soft photons and shape the spectrum of the observed hard X-ray emission and its polarization. We also attempt to estimate the optical depth of the sheath, although this requires simulations that model the interaction between radiation and the plasma, and electron-ion Coulomb coupling throughout the accretion flow; steps toward such simulations have been made by \cite{Liska+22}.

GRMHD simulations of accretion onto black holes including plasma-radiation interaction \citep{Dexter+21, Liska+22} have shown the formation of a truncated inner thicker disk (and an outer thinner disk) with a strong poloidal field (akin to a MAD state) if the disk is initially threaded with a dominant poloidal flux at large radii. Our simulation aims at modeling the properties of the inner regions of such a disk, including the current sheets, potential dissipation regions, and the fluid and electromagnetic field conditions found there. Unlike, e.g., \citet{Liska+22}, our simulations are non-radiative. While radiative cooling will strongly affect the conditions of the dense material accreting near the midplane, the properties of the jet and its sheath, being dominated by magnetic stresses, are conceivably similar between radiative and non-radiative models. 
So, simulations of accretion disks without radiative cooling could reasonably inform us about the magnetically dominated inner regions of the accretion flow during the low/hard state, in particular to understand the formation of current sheets and other dissipation regions found there. Therefore, the results from the simulation presented in this work can potentially capture some characteristics of the low/hard state dynamics and may serve to model the nonthermal X-rays seen during these phases. 

The paper is organized as follows. In \S\ref{sec:GRRMHD_setup}, we detail the numerical setup of the GRRMHD simulations. \S\ref{sec:GRRMHD_results} contains the results, with \S\ref{subsec:general} characterizing the jet sheath, \S\ref{subsec:current_sheets} exploring the properties of the current sheets in the jet sheath, \S\ref{subsec:Poynting_flux} focusing on the power flowing through the jet sheath, \S\ref{subsec:bulk_motions} investigating the properties of the bulk motions in this region, and \S\ref{subsec:tau} checking whether the inferred optical depth in the jet sheath region that we identify as a corona satisfies the requirements to produce the observed X-ray spectra. Finally, we summarize our findings in \S\ref{sec:summary}.

\section{GRRMHD simulation setup} \label{sec:GRRMHD_setup}

\begin{figure}
    \includegraphics[width=0.5\textwidth]{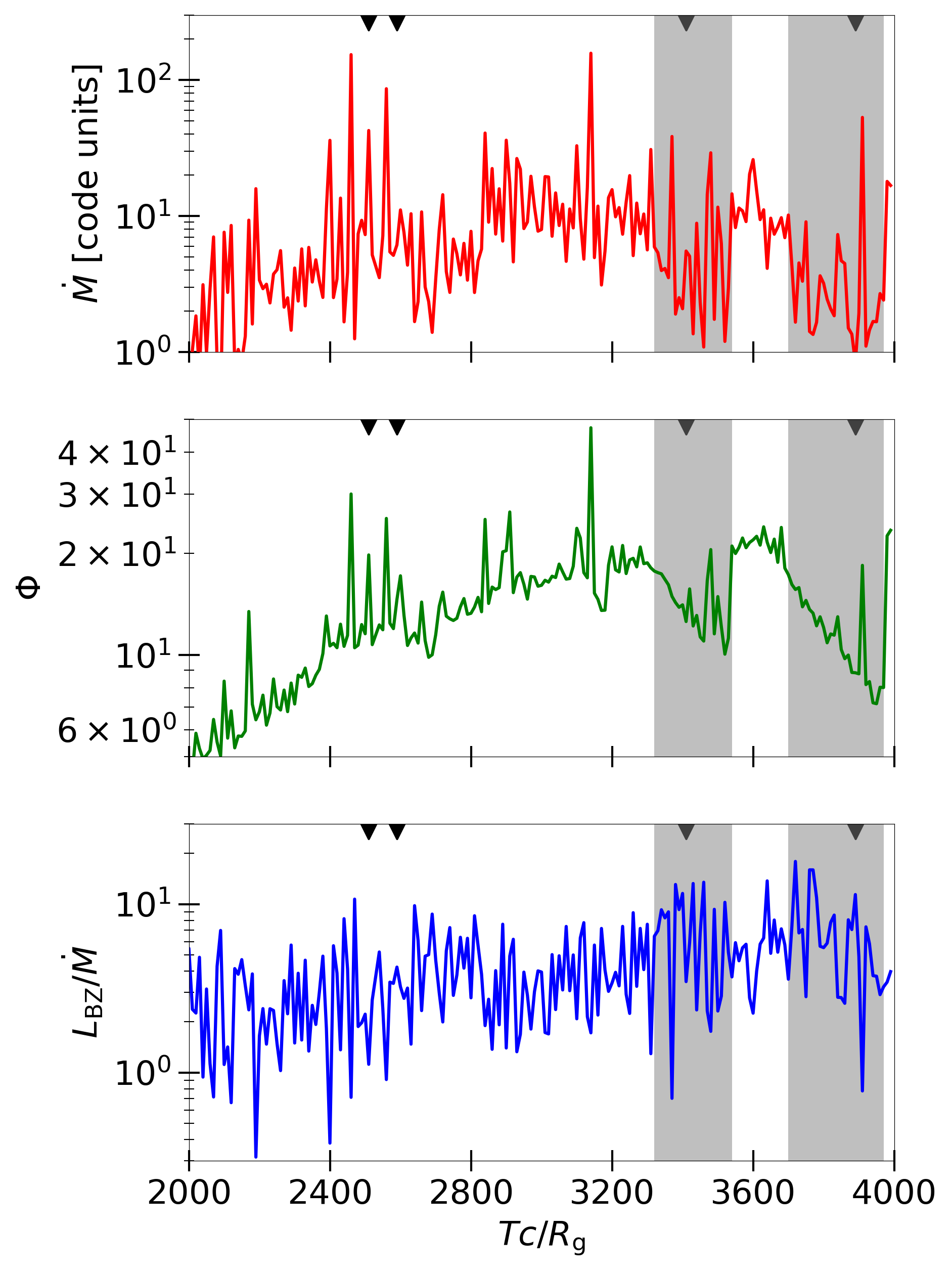}
    \caption{Time evolution of the mass accretion rate $\dot{M}$ (in code units) into the event horizon (top panel), the magnetic flux $\Phi$ threading the horizon (middle panel), and the Blandford-Znajek jet power $L_{\rm BZ}$ normalized by $\dot{M}$ (bottom panel) in our MAD simulation. The downward-facing triangles at the top of each panel denote the times corresponding to the snapshots in Fig.~\ref{fig:x-z_params}. The grey-shaded regions denote phases of post-MAD flux eruptions.}
    \label{fig:horizon_params}
\end{figure}

The results presented in this work are obtained using two-dimensional GRRMHD simulations of prograde accretion onto a highly spinning black hole \citep{Ripperda+20}. The simulation was performed using the Black Hole Accretion Code \citep[{\tt BHAC};][]{Porth+17, Olivares+19} with an implicit-explicit time-stepping scheme \citep{Ripperda+19b} and adaptive mesh refinement (AMR) to accurately resolve dissipation sites on the resistive scale in the system. To accurately capture the regime of plasmoid formation and evolution (which is associated with magnetic dissipation), the Lundquist number $S=L_{\rm cs}v_{\rm A}/\eta$ should be above the threshold of $10^4$ \citep{loureiro_07, bhattacharjee_09, uzdensky_10}. Taking a small, uniform plasma resistivity of $\eta=5\times10^{-5}$ in the GRRMHD equations, since the Alfv\'{e}n speed is $v_{\rm A}/c = \sqrt{\sigma/(\sigma+1)}\sim 1$ in the highly magnetized regions surrounding the jet ($\sigma \gtrsim 1$), the length scale of the resolved current sheets that would go plasmoid-unstable was found to be $L_{\rm cs}\gtrsim{\cal O}(1)\,R_{\rm g}$ \citep{Ripperda+20}, where $R_{\rm g}=GM_\bullet/c^2$ is the gravitational radius of a black hole with mass $M_\bullet$. The simulation results presented here thus reliably capture plasmoid formation in current sheets with length $\gtrsim R_{\rm g}$ near the black hole.

\begin{figure*}
    \includegraphics[width=\textwidth]{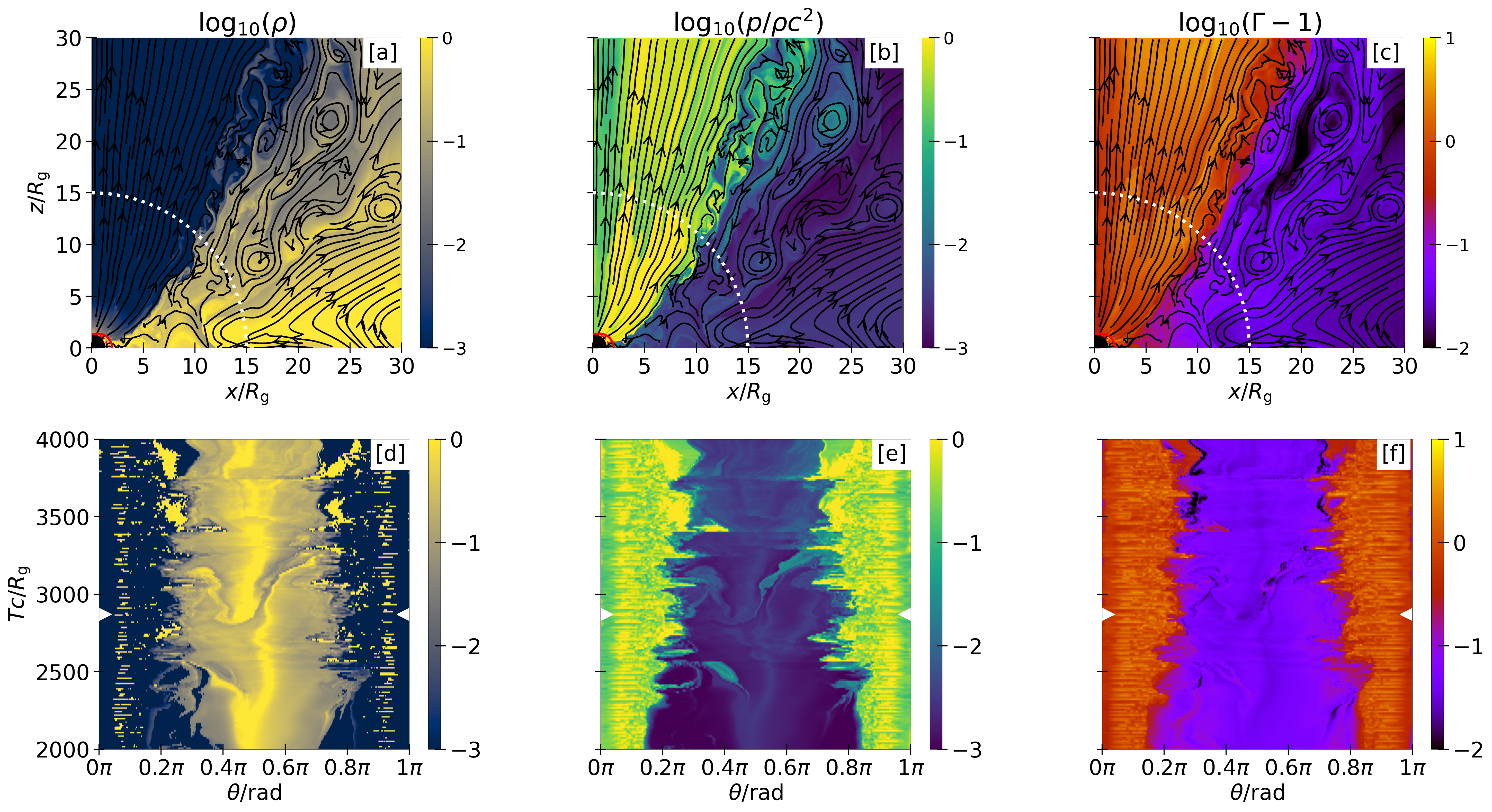}
    \caption{Panels [a,b,c] in the top row show the comoving plasma density, the dimensionless temperature ($p/\rho c^2$), and the bulk energy per unit rest mass energy ($\Gamma-1$), at time $Tc/R_{\rm g}=2870$. The black hole is centered at $[x,z]=[0,0]$, the red solid curve is the ergosphere, and the red dashed curve is the inner light surface. The overplotted curves are magnetic field lines. The time-evolution of these parameters (during the quasi-steady state, $Tc/R_{\rm g}>2000$) at a fiducial radius of $r=15\,R_{\rm g}$ is shown in the bottom row panels [d,e,f] as a function of the polar angle $\theta$ (the white dotted curves in the top row indicate the fiducial radius). The white left- and right-facing triangles in each of the bottom panels denote the time corresponding to the top panels. We only show the northern hemisphere, $\theta<\pi/2$.}
    \label{fig:theta-time_parameters}
\end{figure*}

From hereon, we use the geometrized system of units with gravitational constant, black hole mass, and speed of light $G=M_\bullet=c=1$; length and time scales are thus normalized to $R_{\rm g}$ and $R_{\rm g}/c$, respectively. We also adopt the Lorentz-Heaviside units where a factor of $1/\sqrt{4\pi}$ is absorbed into the electromagnetic fields, and we employ a $3+1$ decomposition of the GRRMHD equations based on the Arnowitt-Deser-Misner formalism \citep{Arnowitt+59} with a $(-,+,+,+)$ signature for the space-time metric. The simulation is performed in a logarithmic Kerr-Schild coordinate system \citep{McKinney&Gammie_04}, with $r, \theta, \phi$ being the radial, polar, and toroidal coordinates (for a logarithmic Kerr-Schild metric, the radial coordinate is $\log r$). The determinant of the metric is $g=-r^2\Sigma^2\sin^2{\theta}$, and the determinant of the spatial component of the metric is
\be
\gamma = g_{\theta\theta}(r,\theta) \left[g_{rr}(r,\theta)g_{\phi\phi}(r,\theta) - g_{r\phi}(r,\theta)^2\right],
\ee
where the spatial metric components are 
\begin{align}
g_{rr}(r,\theta) & = r^2[1 + \zeta(r,\theta)], \nonumber \\
g_{\theta\theta}(r,\theta) & = \Sigma(r,\theta), \nonumber\\
g_{\phi\phi}(r,\theta) & = \left[r^2 + a_\bullet^2 + 2ra_\bullet^2\sin^2{\theta}/\Sigma \right]\sin^2{\theta}, \nonumber \\
g_{r\phi}(r,\theta) & = -ra_\bullet\sin^2{\theta} \left[1+\zeta(r,\theta)\right],
\end{align}
where $\Sigma = r^2 + a_\bullet^2 \cos^2{\theta}$, $\Delta(r)=r^2 + a_\bullet^2 -2r$, and $\zeta(r,\theta)=2r/\Sigma(r,\theta)$.

The base resolution of the simulation is $384\times192$ (the two-dimensional nature of our simulation leaves the plasma dynamics along the $\phi$ coordinate invariant), and the simulation employs five levels of AMR, effectively corresponding to a resolution of 6144 radial cells and 3072 polar cells. The simulation is initiated from a torus in hydrodynamic equilibrium \citep[with an equilibrium pressure distribution $p_{\rm eq}$;][]{Fishbone&Moncrief_76} around a black hole with a dimensionless spin parameter $a_\bullet=0.9375$. The torus's inner radius is at $r_{\rm in}=20\,R_{\rm g}$, and its density peaks at $r_{\rm max}=42\,R_{\rm g}$.  The initiated torus is threaded by a weak poloidal magnetic field loop with a magnetic field strength set such that the plasma beta $\beta=2p/b^2=100$, where $p$ is the plasma pressure, and $b$ is the magnetic field strength in the frame comoving with the fluid \citep{Bucciantini&DelZanna_13}. We also define a `cold magnetization' $\sigma_{\rm c}\equiv b^2/(\rho c^2)$ and a  `hot magnetization' $\sigma_{\rm h}\equiv b^2/(\rho h c^2)$, where $\rho$ is the comoving plasma mass density and $h=1 + 4p/(\rho c^2)$ is the specific enthalpy of an ideal gas with an adiabatic index of $\hat{\gamma}=4/3$. The atmosphere of the disk is permeated with diffuse plasma such that its Lorentz factor $\Gamma<20$, rest-mass density $\rho_{\rm atm}=\rho_{\rm min}r^{-3/2}$, and pressure $p_{\rm atm}=p_{\rm min}r^{-5/2}$, where $\rho_{\rm min}=10^{-4}$ and $p_{\rm min}=10^{-6}/3$ are the floor values of the density and pressure of the gas, respectively. The ceiling on the cold magnetization is set to be $\sigma_{\rm c,max}=100$. The magnetorotational instability \citep{velikhov1959stability, Chandrasekhar_60, Balbus&Hawley_91} is triggered by perturbing the equilibrium fluid pressure as $p=p_{\rm eq}(1+X_{\rm p})$ with a uniformly distributed random variable $X_{\rm p}\in [-0.02,0.02]$.

\section{GRRMHD simulation results} \label{sec:GRRMHD_results}

\begin{figure*}
    \includegraphics[width=0.95\textwidth]{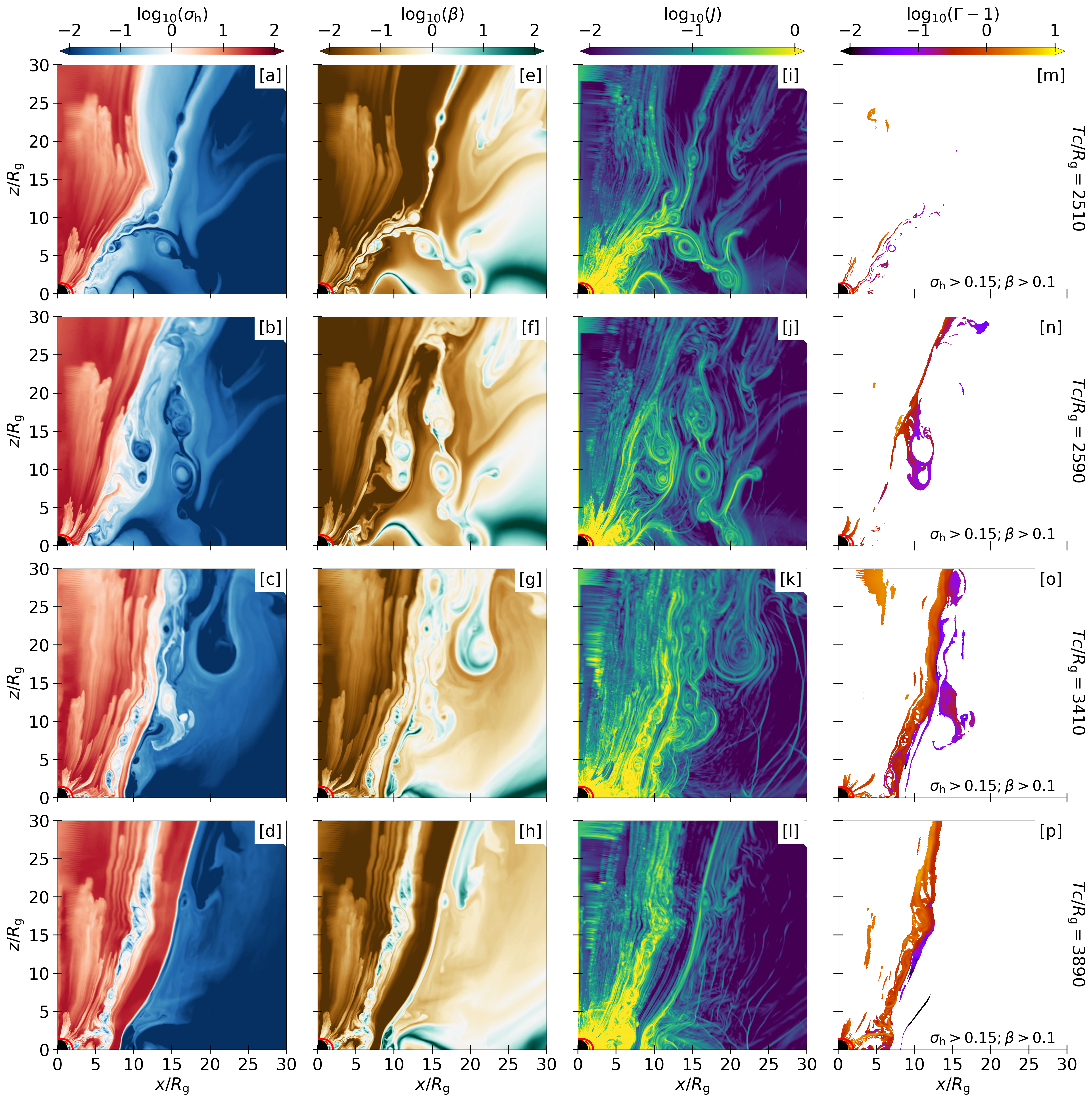}
    \caption{Realizations of plasmoid-mediated reconnection layers identified in the jet sheath by the conditions $\sigma_{\rm h}>0.15$ and $\beta>0.1$ at various times (in different rows). Panels [a-d] show $\sigma_{\rm h}$; panels [e-h] show $\beta$; panels [i-l] show the current density.  The bulk energy per particle in this region (i.e., where $\sigma_{\rm h}>0.15$ and $\beta>0.1$)  is shown in panels [m-p]. In all panels, the black hole is centered at $[x,z]=[0,0]$, the red dashed curve is the ergosphere, and the red solid curve is the inner light surface. Click \href{https://youtu.be/cMmt8FYJI3o}{here} for a YouTube video of some of these quantities.}
    \label{fig:x-z_params}
\end{figure*}

\begin{figure*}
    \includegraphics[width=\textwidth]{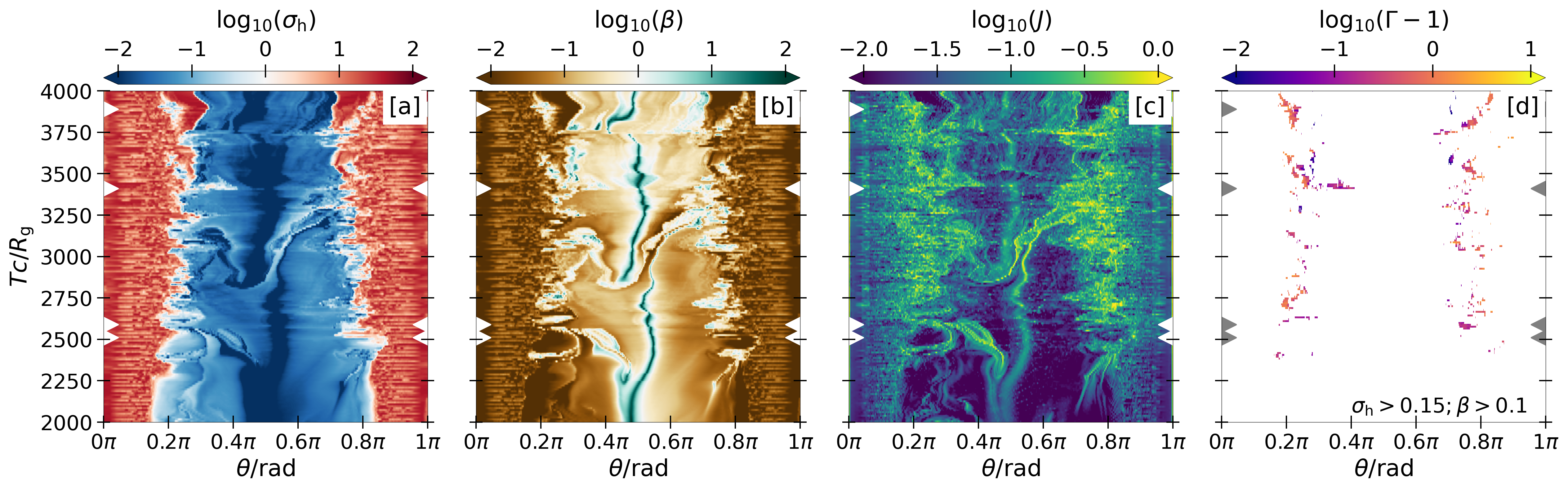}
    \caption{Time-evolution of the quantities in Fig.~\ref{fig:x-z_params} as a function of the polar angle $\theta$, at the fiducial radius $r=15\,R_{\rm g}$. The left- and right-facing triangles denote the times corresponding to different rows of Fig.~\ref{fig:x-z_params}.}
    \label{fig:theta-time_plasmabeta}
\end{figure*}

Fig.~\ref{fig:horizon_params} shows the time evolution of the mass accretion rate (top panel),
\begin{equation} 
\label{eq:mdot}
\dot{M}=\int \rho u^{\rm r} \sqrt{-g}{\rm d}\theta{\rm d}\phi, 
\end{equation}
the magnetic flux threading the event horizon (middle panel), 
\begin{equation}
\Phi=\frac{1}{2}\int\vert B^{\rm r}\vert \sqrt{-g}{\rm d}\theta{\rm d}\phi,    
\end{equation}
and the Blandford-Znajek power of a parabolic jet \citep[lower panel, normalized by $\dot{M}$;][]{Blandford&Znajek_77, Tchekhovskoy+11},
\begin{equation}
L_{\rm BZ}=0.044 (a_\bullet/2)^2 4\pi\Phi^2,    
\end{equation}
where in the above equations, $u^{\rm r}$ and $B^{\rm r}$ are the radial components of the four-velocity and magnetic field, respectively. We define the phase $2000\lesssim Tc/R_{\rm g}\lesssim3000$ when $\dot{M}$ and $\Phi$  increase as the accretion phase. In the rest of the paper, we consider $Tc/R_{\rm g}>2000$ to be the `quasi-steady state' (before which the magnetorotational instability is not well developed), and we perform time averages of various quantities within the range $2000<Tc/R_{\rm g}<4000$. The accretion phase eventually leads to a MAD phase corresponding to saturation in $\Phi$ and $L_{\rm BZ}/\dot{M}$; this is followed by events of flux eruption with a sudden drop in $\Phi$ (gray shaded regions); these phases repeat, although our simulation covers only a few cycles. In the following sections, we describe the properties of the inner regions of the accretion flow, with a specific focus on the sheath at the interface between the jet and the disk.

\subsection{General properties of the jet sheath} \label{subsec:general}

In the top panel of Fig.~\ref{fig:theta-time_parameters}, we present two-dimensional snapshots of the plasma density, the dimensionless temperature ($p/\rho c^2$), and the bulk energy per unit rest mass energy ($\Gamma-1$) at time $Tc/R_{\rm g}=2870$. One can identify the jet sheath as the region where the dimensionless temperature is $p/\rho c^2\sim0.1$ and the bulk energy is $\Gamma-1\sim 1$. In the jet sheath, the flow transitions from being ultra-relativistically fast and hot in the jet to non-relativistic and cold in the disk wind. The plasma mass density is $\rho\sim0.1$ (in code units) at the jet-disk interface. In the following sections, we present a different criterion---based on $\sigma_{\rm h}$ and $\beta$---to identify the regions with reconnection-like features in the jet sheath. 

The jet sheath is also a shear layer that exhibits roll-ups/vortices, produced either by the nonlinear evolution of an in-situ linear instability (e.g., Kelvin-Helmholtz) or by the shearing of large-amplitude waves emanating from near the black hole \citep{Wong+21, Davelaar+23}; such waves could influence nonthermal particle energization and bulk motions in the jet sheath (\citealt{sironi+21a}; more on this in \S\ref{subsec:bulk_motions}). Such roll-ups are also present at this time (e.g., see $[x/R_{\rm g},z/R_{\rm g}]=[10-20,10-20]$). The streamlines overlaid on top of panels [a,b,c] represent magnetic field lines. Note the reversals in the magnetic field direction at the jet sheath: these regions are current sheets, i.e., the jet sheath is both a velocity shear layer and a magnetic shear layer. 

The time evolution of the aforementioned quantities ($\rho, p/\rho c^2, \Gamma-1$) as a function of the polar angle $\theta$ at a fiducial radius of $15\,R_{\rm g}$ is shown in the bottom row. We choose this as the fiducial radius in this paper because reverberation studies have suggested that the nonthermal X-rays emerge from a region within $\sim15\,R_{\rm g}$ from the black hole \citep{Alston+20}. At this radius, the jet sheath is seen to be located at $\theta\sim0.2\pi$ in the northern hemisphere (and symmetrically, at $0.8\pi$ in the southern hemisphere) during the quasi-steady state. 
The bright horizontal streaks denoting increased density (in panel [d]) and temperature (in panel [e]) seen within the jet core ($\theta\lesssim0.1\pi$ and $\gtrsim 0.9\pi$) are numerical artifacts, where the code injects additional plasma since the density (or magnetization) hits its floor (ceiling) value (see \S\ref{sec:GRRMHD_setup}).

\subsection{Characteristics of the reconnection layers} \label{subsec:current_sheets}
\subsubsection{Location and geometry} 

Current sheets of various lengths form in the jet sheath, which can result in magnetic reconnection. In Fig.~\ref{fig:x-z_params}, we show the current sheets that form close to the black hole at various times ($Tc/R_{\rm g}=2510,\,2590,\,3410,\,3890$; top to bottom rows).  The current sheets/reconnection layers are visualized by the following parameters: $\sigma_{\rm h}$ (first column), plasma $\beta$ (second column), and current density magnitude $\sqrt{J^iJ_i}$ (third column), where 
\be
J^i = qv^i + \frac{\Gamma}{\eta} \left[E^i + \gamma^{-1/2}\varepsilon^{ijk}\frac{v_j}{c}B_k - (v^jE_j)\frac{v^i}{c^2}\right],
\ee
with $q=\nabla \cdot \mathbf{E}$ being the charge density, $E^{i}$ and $B^{i}$ being the electric and magnetic field three-vectors, $\varepsilon^{ijk}$ being the Levi-Civita anti-symmetric tensor, and $v^{i}$ being the three-velocity with corresponding bulk Lorentz factor $\Gamma = \sqrt{1 / (1 - v^i v_i/c^2)}$. 

We see large-scale current sheets, with lengths of ${\cal O}(10\,R_{\rm g}$) and widths of ${\cal O}(0.1\,R_{\rm g}$). The current sheets in the jet sheath may be classified depending on whether their magnetic footpoints are rooted (1) on the event horizon of the black hole and the accretion disk or wind (e.g., the arc-like sheet at $Tc/R_{\rm g}=2510$), (2) only the accretion disk (e.g., the nearly vertical sheets at $Tc/R_{\rm g}=2590, 3410$). Case (1) is primarily responsible for the bright `fingers' observed as protruding into $\theta>0.3\pi$ and $\theta<0.8\pi$ in panels [a-b] of Fig.~\ref{fig:theta-time_plasmabeta} \citep[see also][]{El_Mellah+23}.

Note that the presence of a current sheet does not guarantee fast dissipation. However, a current sheet showing plasmoid chains is undergoing active reconnection, which implies fast dissipation. 
Most of the time, long current sheets in the jet sheath are seen to harbor chains of plasmoids, which appear as prominent regions with $\beta\gtrsim1$. Most of the plasmoids residing in/near the sheath at a given time were born in the sheath itself. However,  there are brief moments (of duration $\sim10\,R_{\rm g}/c$) when plasmoids are not formed in situ in the sheath: these are the times immediately following a flux eruption. During those moments, the region adjacent to the sheath (toward the disk, with $\sigma_{\rm h}<0.15$) may still harbor large plasmoids ($\sim5\,R_{\rm g}/c$), which were created at the base of the jet during the prior flux eruption. These plasmoids can grow to large sizes (${\cal O}(10)\,R_{\rm g}$) while propagating outwards. Some examples appear in the top panel of Fig.~\ref{fig:theta-time_parameters} as the closed loops of magnetic field lines (e.g., at $[x/R_{\rm g},z/R_{\rm g}] = [16,7], [24,22]$). Numerical simulations have revealed that some of these large plasmoids move along a spiral trajectory around the jet \citep{Nathanail+20, Ripperda+22, El_Mellah+23}. 

Given the copious presence of plasmoids and reconnecting layers in the jet sheath, we regard this interface as an important site of energy dissipation. As such, it may be a potential candidate for the Comptonizing corona. More quantitatively, we identify the reconnection-powered dissipation regions in the jet sheath by the following combination: plasma beta $\beta>1$ and hot magnetization $\sigma_{\rm h}>0.15$. The former threshold excludes the jet core (which has lower plasma beta, and is unphysically affected by numerical density and pressure floors), while the latter excludes the weakly magnetized disk. In the following, we call the region identified by the two conditions the `dissipative jet sheath,' or the `candidate corona.' 

The regions satisfying our cuts in plasma beta and magnetization are colored by the local bulk Lorentz factor in the last column of Fig.~\ref{fig:x-z_params}. This shows that large plasmoids, reaching sizes $\sim5\,R_{\rm g}$ as a result of mergers with other plasmoids, can have a range of velocities, depending on their proximity to the jet sheath. Plasmoids close to the jet sheath are trans-relativistic, with $\Gamma-1\sim1$ (e.g., at $[x/R_{\rm g},z/R_{\rm g}]\sim[10,10]$ in the third row of Fig.~\ref{fig:x-z_params}, or at $[x/R_{\rm g},z/R_{\rm g}]=[16,20]$ in Fig.~\ref{fig:theta-time_parameters}[c]), while plasmoids that are closer to the disk are sub-relativistic, with $\Gamma-1\sim10^{-2}$ (e.g., at $[x/R_{\rm g},z/R_{\rm g}]\sim[12,13]$ in the second row of Fig.~\ref{fig:x-z_params}, or at $[x/R_{\rm g},z/R_{\rm g}]=[24,22]$ in Fig.~\ref{fig:theta-time_parameters}[c]). As we further discuss in \S\ref{subsec:bulk_motions}, their motion is largely determined by global stresses, rather than by the local reconnection dynamics (i.e., the release of magnetic tension of the reconnected fields).

As shown in Fig.~\ref{fig:x-z_params}, the thickness of the dissipative jet sheath can vary significantly over time. We find that it is wider during flux eruption phases (grey regions in Fig.~\ref{fig:horizon_params}), when $\Phi$ drops to a minimum at the event horizon. In contrast, it gets thinner---and nearly devoid of large plasmoids---during the MAD phase, when flux accumulation has saturated at the event horizon, i.e., at times just before flux eruptions. In both cases, typical bulk motions in the dissipative jet sheath are trans-relativistic, see the last column of Fig.~\ref{fig:x-z_params}.

The time-evolution of $\sigma_{\rm h}$, $\beta$, and $J$ at the fiducial radius of $15\,R_{\rm g}$ is shown in Fig.~\ref{fig:theta-time_plasmabeta} as a function of the polar angle $\theta$. This highlights the copious presence of regions with large current density located at the jet sheath ($\theta\sim0.2\pi, 0.8\pi$) at different times during the quasi-steady state, $Tc/R_{\rm g}>2000$. The presence of trans-relativistic bulk motions in this region (also identified by the $\sigma_{\rm h}>0.15, \beta>0.1$ cuts) is seen to recur on a timescale of ${\cal O}(10-100)\,R_{\rm g}/c$ (panel [d]): this corresponds to a frequency of $\sim100-1000$\,Hz for a $10\,M_{\rm \odot}$ black hole. Therefore, photons that diffuse across a $10-100\,R_{\rm g}$ region with an optical depth of order unity will always encounter some realization of the reconnecting layers in the jet sheath.

\subsubsection{Azimuthal field strength in the reconnection layers} \label{subsubsec:guide_field}

\begin{figure*}
    \includegraphics[width=\textwidth]{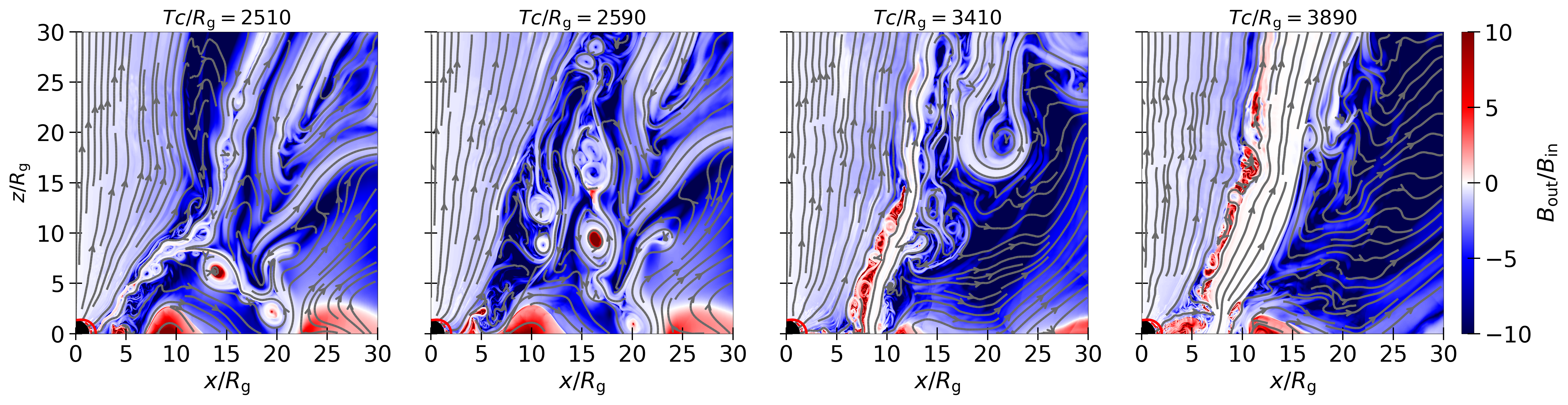}
    \caption{Strength of the out-of-plane magnetic field $B_{\rm out}$ at different times (same times as in Fig.~\ref{fig:x-z_params}), normalized to the strength of the in-plane field $B_{\rm in}$. The streamlines denote in-plane magnetic field lines.}
    \label{fig:guide_field}
\end{figure*}

\begin{figure*}
    \includegraphics[width=\textwidth]{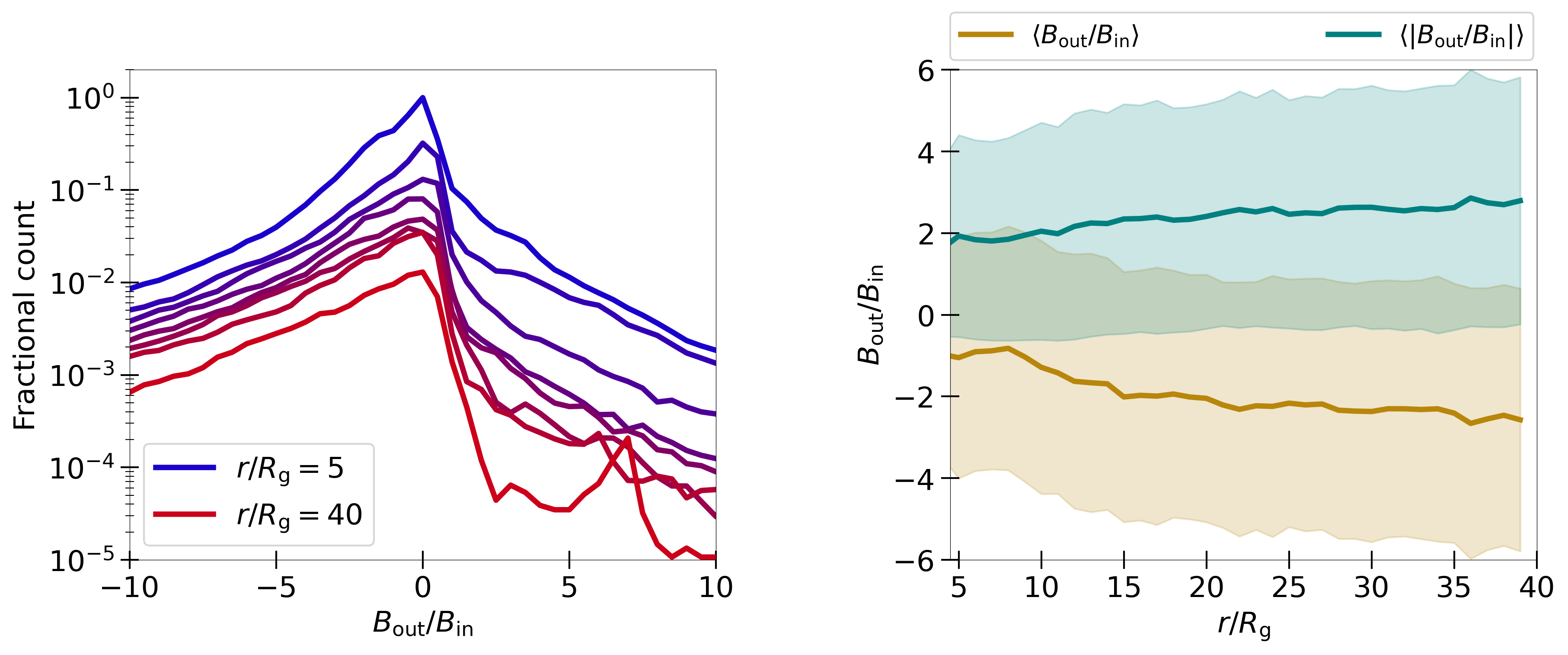}
    \caption{Left panel: time- and $\theta$-averaged histogram of the out-of-plane field strength ($B_{\rm out}$) in units of the in-plane field strength ($B_{\rm in}$), measured between 5\,$R_{\rm g}$ (blue) and 40\,$R_{\rm g}$ (red) in increments of 5\,$R_{\rm g}$. The histograms are normalized with respect to the peak of the histogram for the $r=5\,R_{\rm g}$ case. Right panel: radial dependence of the average (solid curve) and standard deviation (shaded region) of $B_{\rm out}/B_{\rm in}$ (brown) and $|B_{\rm out}/B_{\rm in}|$ (teal). All quantities presented in this figure are calculated within the dissipative jet sheath (identified by the cuts $\sigma_{\rm h}>0.15$ and $\beta>0.1$) in the upper hemisphere, and averaged over the quasi-steady state, $2000<Tc/R_{\rm g}<4000$.}
    \label{fig:guide_field_histogram}
\end{figure*}

\begin{figure*}
    \includegraphics[width=\textwidth]{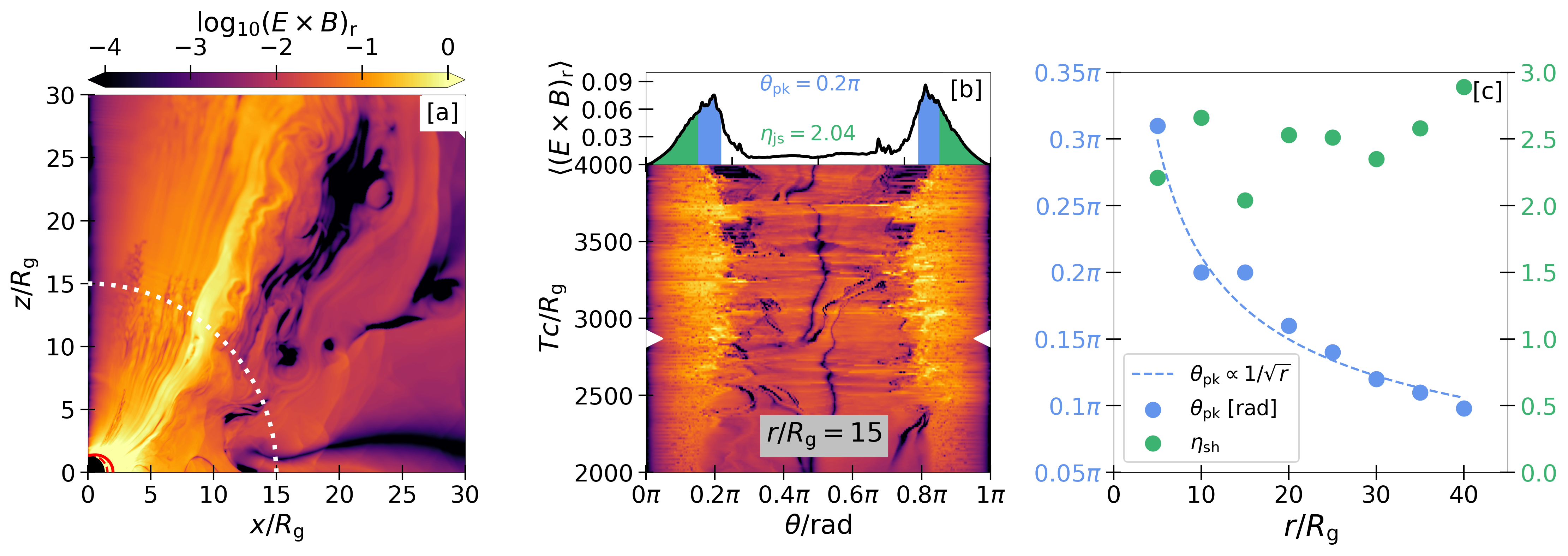}
    \caption{Panel [a]: Radial Poynting flux at time $Tc/R_{\rm g}=2870$. Panel [b]: Time-evolution of the radial Poynting flux as a function of the polar angle $\theta$ at the fiducial radius of $r=15\,R_{\rm g}$. The black curve in the top sub-figure shows the time-average $\langle(E\times B)_{\rm r}\rangle[\theta]$. The angle where $\langle(E\times B)_{\rm r}\rangle$ peaks, $\theta_{\rm pk}$, is mentioned in the top sub-panel (blue text). The blue area in the top sub-panel defines the jet sheath such that $\langle(E\times B)_{\rm r}\rangle[\theta] \ge 0.75\langle(E\times B)_{\rm r}\rangle[\theta_{\rm pk}]$. The green area is the jet core. The ratio of the radially-flowing electromagnetic power in the jet sheath to that in the jet core is denoted by $\eta_{\rm js}$ (green text). Panel [c]: Blue and green markers show $\theta_{\rm pk}$ and $\eta_{\rm js}$ as a function of radius (corresponding to the left and right vertical axes). The blue dashed curve is an empirical fit to $\theta_{\rm pk}(r)$, scaling as $\propto 1/\sqrt{r}.$} 
    \label{fig:ExBr_theta_time_f_csjet}
\end{figure*}

Fig.~\ref{fig:guide_field} shows the magnetic field structure at different times (same times as in Fig.~\ref{fig:x-z_params}). Streamlines display the in-plane magnetic field lines, demonstrating that it is primarily the radial component (and not the polar component) that undergoes reconnection.
Colors in Fig.~\ref{fig:guide_field} denote the strength of the out-of-plane (azimuthal) field in the Eulerian frame \citep[see discussion around Eq.~1 of][for the definition]{Ripperda+19a}, normalized by the in-plane (poloidal) field, 
\be \label{eq:B_out}
B_{\rm out}/B_{\rm in} = \sqrt{g_{\phi\phi}}B^{\phi}/(g_{rr}B^{r} B^{r} + g_{\theta\theta}B^\theta B^\theta)^{1/2}.
\ee
The figure shows that the azimuthal field in the northern hemisphere is preferentially negative (the opposite holds in the southern hemisphere), although some dissipation regions in the jet sheath have mostly positive $B_{\rm out}$ (see third and fourth panels). During flaring events, magnetic loops originating in the southern hemisphere (and thus preferentially having $B_{\rm out}>0$) may be launched into the jet sheath of the upper hemisphere, as illustrated by the third and fourth panels in Fig.~\ref{fig:guide_field}. We also point out that at the equatorial current sheet appearing during flux eruptions, all field components (i.e., both in-plane fields and $B_{\rm out}$) flip their polarity \citep{Ripperda+20,Ripperda+22}, i.e., here reconnection proceeds in the regime of vanishing guide field.

We now characterize in more detail the magnetic field geometry in the dissipative jet sheath, identified via the thresholds $\beta>0.1$ and $\sigma_{\rm h}>0.15$ as described above. A time- and $\theta$-averaged histogram of $B_{\rm out}/B_{\rm in}$ in the dissipative jet sheath is shown in the left panel of Fig.~\ref{fig:guide_field_histogram}. We consider only the upper hemisphere (opposite results hold for the lower hemisphere), and we average over the quasi-steady state.
The radial dependence is captured by curves with different colors, from 5\,$R_{\rm g}$ (blue) to 40\,$R_{\rm g}$ (red) in increments of 5\,$R_{\rm g}$. The preponderance of negative azimuthal fields in the upper hemisphere (see the blue regions in Fig.~\ref{fig:guide_field}) drives the asymmetry in the histograms. Note that the histograms are more symmetric (around $B_{\rm out}/B_{\rm in}=0$) close to the black hole (blue curves) than further away (red curves). This is related to the `contamination' by $B_{\rm out}>0$ loops that originated in the lower hemisphere, as described above for the third and fourth panels of  Fig.~\ref{fig:guide_field}. These `contaminating' $B_{\rm out}>0$ regions do not extend far from the black hole, which explains why the histograms become more asymmetric at larger distances.

Using these histograms, we find the time-averaged azimuthal field in the reconnection regions of the dissipative jet sheath (i.e., where $\beta>0.1$ and $\sigma_{\rm h}>0.15$).
It decreases from $\langle B_{\rm out}/B_{\rm in}\rangle \pm\sigma_{B_{\rm out}/B_{\rm in}}=-1\pm3$ at $r=5\,R_{\rm g}$ to $-2.6\pm3.2$ at $r=40\,R_{\rm g}$.\footnote{From hereon, we denote the standard deviation of a quantity $X$ as $\sigma_{X}$.} The mean magnitude of the azimuthal field   
increases from $\langle|B_{\rm out}/B_{\rm in}|\rangle \pm\sigma_{|B_{\rm out}/B_{\rm in}|}=1.9\pm2.5$ at $r=5\,R_{\rm g}$ to $2.8\pm3.0$ at $r=40\,R_{\rm g}$. 

If the in-plane field reverses polarity across a current sheet, while the out-of-plane component does not, then one could interpret the out-of-plane component---along the direction of the electric current---as a guide field. 
While $B_{\rm out}$ does not always act as a guide field in the dissipative jet sheath, it could locally act as a guide field at certain times, as its polarity often stays constant across the current sheets in the jet sheath (e.g., see first and second panels of Fig.~\ref{fig:guide_field}).\footnote{This is true in our two-dimensional simulation and may change in three-dimensional simulations where physical quantities will not be invariant in the $\phi$ direction.} 
In contrast, as already anticipated above, across the equatorial current sheet all field components reverse in polarity, so the out-of-plane azimuthal field cannot be interpreted as a guide field there.
The properties of particle energization, reconnection plasmoids, and their bulk motions depend on the strength of the out-of-plane guide field, as demonstrated by local kinetic studies of reconnection \citep[e.g.,][]{ball_18, Gupta+24}. Current sheets that undergo reconnection in the presence of large guide fields in the jet sheath might also be sites of proton acceleration that can produce high-energy neutrinos \citep{Fiorillo+23}.  

\subsection{Power in the jet sheath} \label{subsec:Poynting_flux}

The power flowing along the jet sheath can be reasonably estimated considering only the electromagnetic component, since the sheath is magnetically dominated, $\sigma_{\rm h}\gtrsim1$ (see left column of Fig.~\ref{fig:x-z_params}). The dominant component of the  Poynting flux is the radial one\footnote{See Appendix~\ref{Appendix:Poynting_components} for comparison to polar and azimuthal components.}. Its covariant form \footnote{Note that the physical component of radial Poynting flux in the orthonormal basis is $\sqrt{g^{\rm rr}} (E\times B)_{\rm r}$.} 
is given by,
\be
(E\times B)_{\rm r} = \sqrt{\gamma} \left(E^\theta B^\phi - E^\phi B^\theta\right).
\ee
Fig.~\ref{fig:ExBr_theta_time_f_csjet} shows the covariant Eulerian $(E\times B)_{\rm r}$ at $Tc/R_{\rm g}=2870$ in panel [a]. We show the time evolution of $(E\times B)_{\rm r}$ as a function of the polar angle $\theta$ in the bottom sub-figure of panel [b], for our fiducial radius $r=15\,R_{\rm g}$. The top sub-figure shows the time-averaged $\langle(E\times B)_{\rm r}\rangle$ as a function of $\theta$. The polar angle $\theta_{\rm pk}$ where  $\langle(E\times B)_{\rm r}\rangle$ peaks  is $\sim0.2\pi$ for the northern hemisphere (and symmetrically, $\sim0.8\pi$ for the southern hemisphere) at $r=15\,R_{\rm g}$. We identify the angular extent of the jet sheath as the range where $\langle(E\times B)_{\rm r}\rangle[\theta]\ge0.75\langle(E\times B)_{\rm r}\rangle[\theta_{\rm pk}]$ (denoted by blue regions in Fig.~\ref{fig:ExBr_theta_time_radii}). The green regions ($\lesssim0.2\pi$ and $\gtrsim0.8\pi$)  identify the jet core.
The factor of 0.75 that we adopt yields an angular width that is comparable to the width of the region identified by the $\sigma_{\rm h}>0.15$ and $\beta>0.1$ thresholds (more on this in Appendix~\ref{Appendix:Poynting_efficiency}). In panel [c], we show that $\theta_{\rm pk}$ (measured in the northern hemisphere) decreases from $\sim0.3\pi$ at a radius of $5\,R_{\rm g}$ to $\sim0.1\pi$ at a radius of $40\,R_{\rm g}$. This can be fitted with a function of the form $\theta_{\rm pk}\simeq 0.3\pi(r/5\,R_{\rm g})^{-0.5}$, indicating a paraboloidal shape. 

The ratio of the (polar-angle-integrated) time-averaged electromagnetic power in the jet sheath (blue region) to that in the jet core  (green region) is found to be $\eta_{\rm js}\simeq2$ at $15\,R_{\rm g}$ (in Appendix~\ref{Appendix:Poynting_efficiency}, we compute $\eta_{\rm js}$ adopting different definitions). As shown by the green dots in Fig.~\ref{fig:ExBr_theta_time_f_csjet}[c] (and in the top sub-figures of Fig.~\ref{fig:ExBr_theta_time_radii}), this ratio stays at $\eta_{\rm js}\sim2-3$ regardless of the distance from the black hole (at least until $40\,R_{\rm g}$). 
For a highly spinning black hole, the jet efficiency, defined as the ratio of power flowing in the jet core (where $\beta<1$) to the accretion power $\dot{M}c^2$, is measured to be $\lesssim50\%$ from radiative GRMHD simulations \citep{Liska+22} intended to represent hard intermediate states (i.e., different from the non-radiative thick disk MAD simulated here). 
Since the jet sheath carries twice as much power as the jet core, the power carried by the dissipative jet sheath is comparable to the accretion power. This does not violate energy conservation, since most of the jet power (both jet core and jet sheath) comes from extraction of the black hole spin energy. 

The above findings are useful for interpreting the hard state observations of black hole X-ray binaries \citep{McClintock&Remillard_06}. For instance, the nonthermal X-ray luminosity of Cygnus X-1 during the hard state is $\sim10^{37}\,{\rm erg\,s^{-1}}$, which is comparable to its jet power, as measured directly from the interaction of the jet with the surrounding medium using  H$\alpha$ and [O\,{\sc iii}] measurements of the jet-powered nebula \citep{Gallo+05, Russell+07}. The fact that the electromagnetic power flowing in the jet sheath is about twice that compared to the power flowing in the jet core suggests that the jet sheath might carry enough energy to power the observed nonthermal X-ray emission. 

In an attempt to quantify the dissipation occurring in the jet sheath, we also calculate (in Appendix~\ref{Appendix:dissipation}) the radial dependence of the radially-outflowing electromagnetic power, 
\be
\dot{E}^{\rm (EM)}(r) = 2\pi\int\sqrt{g^{\rm rr}} (E\times B)_{\rm r}\sqrt{-g}{\rm d}\theta.
\ee 
We find that along the jet sheath, $\dot{E}^{\rm (EM)}(r)$ decreases with radius as $\propto r^{-0.2}$ from $2-10\,R_{\rm g}$ (demonstrating magnetic dissipation) then increases with radius (shown in Fig.~\ref{fig:ExBr_radial_corona} until $\sim40\,R_{\rm g}$): note that an increasing $\dot{E}^{\rm (EM)}(r)$ at larger radii does not mean there is no dissipation there; instead, it could be due to the additional contributions from disk winds \citep[as discussed in][]{Qian+18}. For the parameters typical of Cygnus X-1 (mentioned below in \S\ref{subsec:tau}), we find that up to $\sim10^{38}\,{\rm erg\,s^{-1}}$ may be dissipated along the jet sheath from $2-10\,R_{\rm g}$.

\subsection{Bulk motions in the dissipative jet sheath} \label{subsec:bulk_motions}

\begin{figure*}
    \includegraphics[width=\linewidth]{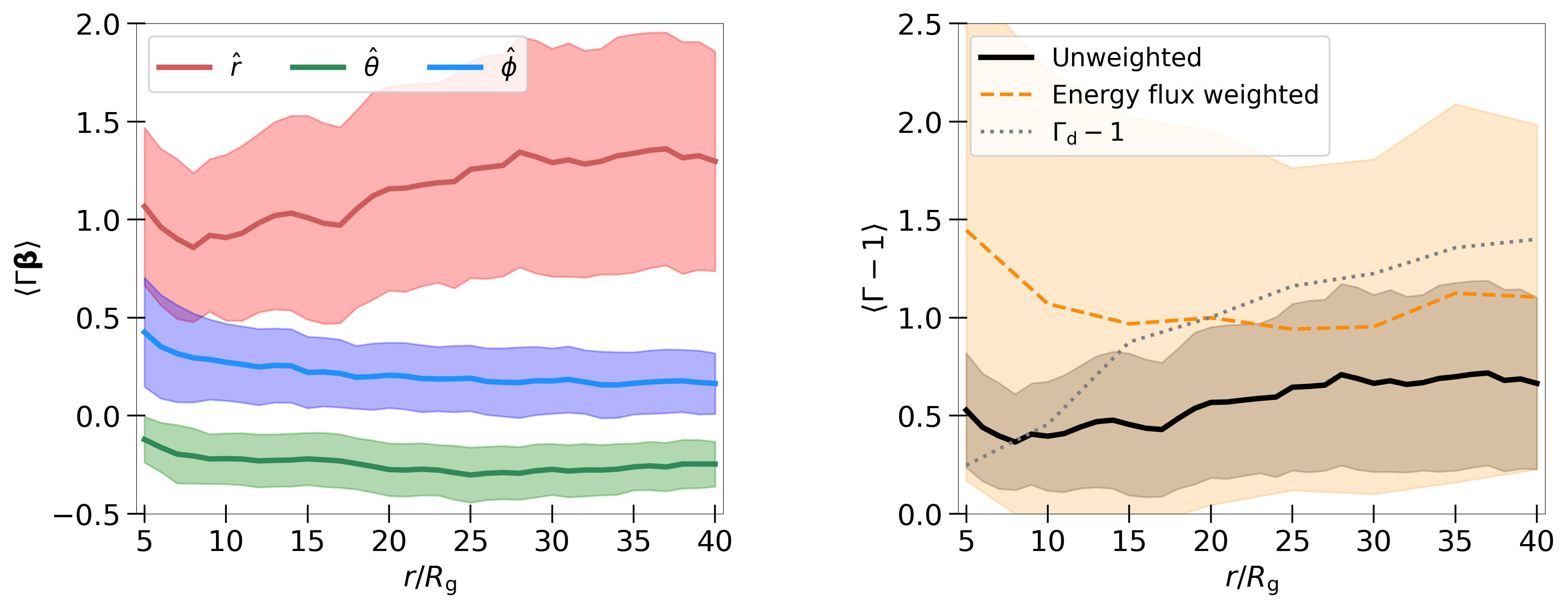}
    \caption{Left panel: 
    solid curves denote the radial dependence of the average of the 4-velocity components $\Gamma\beta_{\rm r}$ (red), $\Gamma\beta_{\rm \theta}$ (green), and $\Gamma\beta_{\rm \phi}$ (blue). Right panel: the black solid curve shows the mean bulk energy $\langle\Gamma-1\rangle$ as a function of radius. The dashed yellow curve shows the mean bulk energy weighted by the radial energy flux $(E\times B)_{\rm r}(1 + 1/\sigma_{\rm h})$. The black dotted curve denotes the bulk Lorentz factor at the jet sheath (assuming $\theta=\theta_{\rm pk}$) expected from $E\times B$ drift motion, see (Eq.~\ref{eq:Gamma_d1}). The shaded region around each curve denotes the standard deviation. All quantities are measured in the Eulerian frame within the dissipative jet sheath in the upper hemisphere and averaged over the quasi-steady state.}
    \label{fig:coronal_region_bulk_components}
\end{figure*}

\begin{figure*}
    \includegraphics[width=\linewidth]{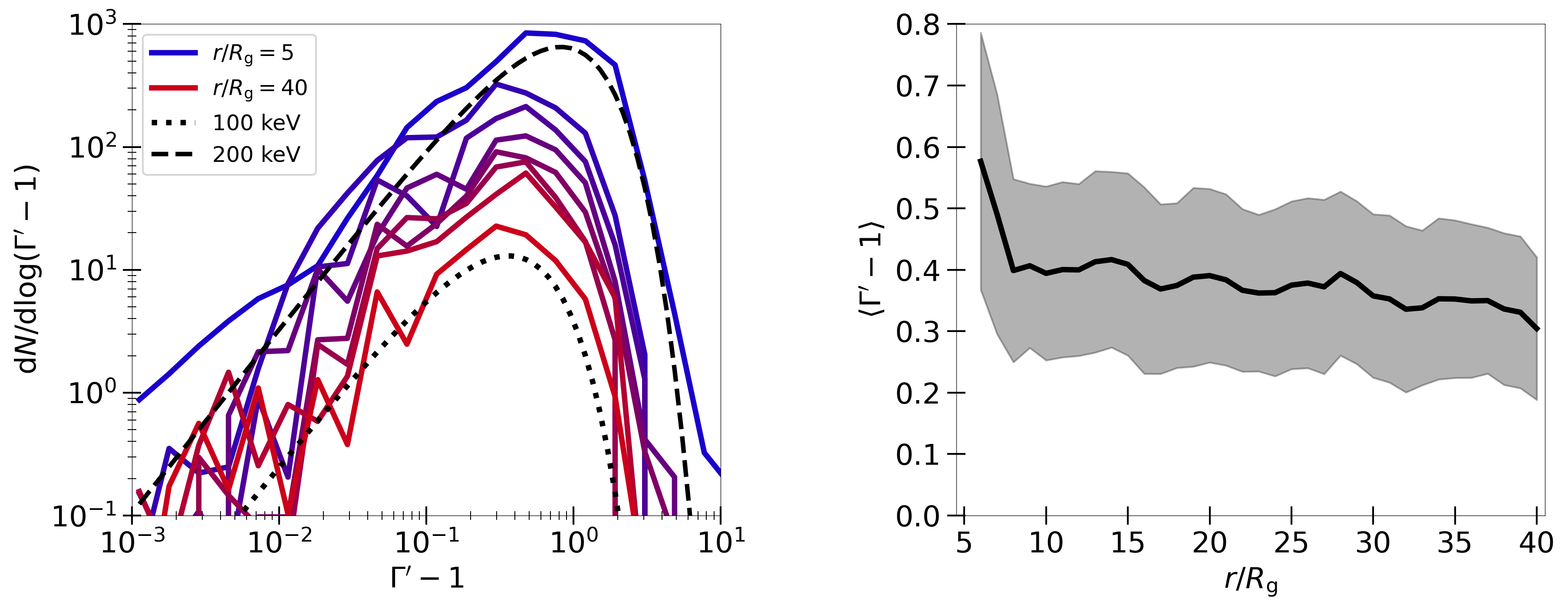}
    \caption{Left panel: time- and $\theta$-averaged bulk energy spectra, calculated in the local comoving frame between 5\,$R_{\rm g}$ (blue) and 40\,$R_{\rm g}$ (red), in bins of  5\,$R_{\rm g}$. The black dotted and dashed curves denote 100\,keV and 200\,keV Maxwellian distributions, respectively. Right panel: mean bulk energy in the comoving frame (solid line) and its standard deviation (shaded band) as a function of radius within the dissipative jet sheath. All quantities are measured in the frame comoving with the fluid within the dissipative jet sheath in the upper hemisphere and averaged over the quasi-steady state.
    }
    \label{fig:coronal_region_bulk_motions}
\end{figure*}

In the left panel of Fig.~\ref{fig:coronal_region_bulk_components}, we show the radial ($\hat{r}$; red), polar ($\hat{\theta}$; green), and azimuthal ($\hat{\phi}$; blue) components of the dimensionless Eulerian bulk 4-velocity ($\Gamma\boldsymbol{\beta}$) as a function of radius. All quantities are averaged over the polar angle within the dissipative jet sheath and in time during the quasi-steady state. The colored shaded band around each curve denotes the corresponding standard deviation. We find that the fluid bulk motions are dominated by the radial component ($\Gamma\beta_{\rm r}$), with $\langle \Gamma\beta_{\rm r} \rangle \pm \sigma_{\Gamma\beta_{\rm r}}$, which increases from $1.0\pm0.4$ at $r=5\,R_{\rm g}$ until it saturates at $1.3\pm0.6$ for $r\gtrsim 30\,R_{\rm g}$. The polar component decreases from $\langle \Gamma\beta_{\rm \theta} \rangle \pm \sigma_{\Gamma\beta_{\rm \theta}} = -0.12 \pm 0.11$ at $r=5\,R_{\rm g}$ to $-0.24\pm0.11$ at $r=40\,R_{\rm g}$. The azimuthal component also decreases with radius, from $\langle \Gamma\beta_{\rm \phi} \rangle \pm \sigma_{\Gamma\beta_{\rm \phi}} = 0.43 \pm 0.28$ at $r=5\,R_{\rm g}$ to $0.16\pm0.15$ at $r=40\,R_{\rm g}$.  The mean bulk motion of the dissipative jet sheath, as quantified by the left panel of  Fig.~\ref{fig:coronal_region_bulk_components}, is crucial in determining (\textit{i}) the energy density and angular distribution of disk photons getting reprocessed in the corona (e.g., \citealt{ Zdziarski+99}), and (\textit{ii}) the fraction of Comptonized X-rays that return to the disk and give rise to the reflected component (e.g., \citealt{Beloborodov_99}). Our result of a radially-accelerating coronal region is in agreement with the MHD models of \cite{Merloni&Fabian_02} and with the empirical model of \cite{King+17}.

As illustrated by the left panel of Fig.~\ref{fig:coronal_region_bulk_components}, the typical bulk motions in the dissipative jet sheath are trans-relativistic. The mean bulk Lorentz factor in the lab frame $\langle \Gamma-1\rangle\pm\sigma_{ \Gamma-1}$ increases from $0.5\pm0.3$ at $r=5\,R_{\rm g}$ until it saturates at $0.66\pm0.44$ for $r\gtrsim 30\,R_{\rm g}$. This is shown by the black solid curve in the right panel of Fig.~\ref{fig:coronal_region_bulk_components}. For comparison, we also show the mean and standard deviation of the bulk Lorentz factor weighted by the radial energy flux $(E\times B)_{\rm r}(1+1/\sigma_{\rm h})$ (dashed yellow curve and shaded region). Here, the factor of $1/\sigma_{\rm h}$ accounts for the contribution of the matter energy flux, assuming that the enthalpy density is dominated by hot (non-radiative) ions. On the other hand, if the dissipative jet sheath were to be dominated by radiatively-cooled electron-positron pairs, one would need to replace the hot magnetization $\sigma_{\rm h}$ with the cold magnetization $\sigma_{\rm c}$. As further discussed below, a proper characterization of the matter composition of the jet sheath requires dedicated radiative simulations.

What is the origin of the trans-relativistic motions in the jet sheath? Given that most of the structures we observe there are related to reconnection, one might speculate that the properties of the bulk motions could be, in part, attributed to the release of magnetic tension of the reconnected field lines and to the global dynamics of the system \citep{Beskin+04}. A lower limit on the fluid speed is given by 
\be \label{eq:v_d}
\beta_{\rm d} = \left|\frac{{\bf E}\times {\bf B}}{B^2}\right| 
\ee
where the magnetic field, ${\bf B} = {\bf B_{\rm p}} + B_{\rm out}\hat{\phi}$, is composed of poloidal and toroidal components. We neglect the component of the electric field parallel to the magnetic field. In the Eulerian frame, the strength of the electric field perpendicular to the magnetic field is given by \citep{Narayan+07}
\be \label{eq:E}
E = \left|-\frac{\Omega R}{c}\hat{\phi} \times {\bf B}\right| = \frac{\Omega r\sin{\theta}}{c} {B_{\rm p}} ,
\ee
where $R=r\sin{\theta}$ is the cylindrical radius. Let us take the angular frequency of the field lines threading the black hole ($\Omega$) to be half of the rotation frequency of the black hole horizon, with radius $r_{\rm H}=(1 + \sqrt{1-a_\bullet^2})R_{\rm g}$ \citep{McKinney&Narayan_07a, McKinney&Narayan_07b}, yielding
\be
\Omega = \frac{a_\bullet c}{4r_{\rm H}} \approx 0.18\,\frac{c}{R_{\rm g}}
\ee 
for $a_\bullet=0.9375$. At the jet boundary ($\theta\sim\theta_{\rm pk}$), far away from the black hole, we have $\Omega r \sin\theta_{\rm pk}\gg 1$, which implies $E\gg B_{\rm p}$. On the other hand, $E\simeq B_{\rm out}$ at most. This implies that the drift Lorentz factor 
$\Gamma_{\rm d}$ can be calculated as
\be \label{eq:Gamma_d1}
{\Gamma_{\rm d}} \!=\!\sqrt{ \frac{B^2 }{B^2- E^2}}\! =\! \sqrt{1\!+\! \frac{B_{\rm out}^2}{B_{\rm p}^2}} \approx \sqrt{1\!+\!\left[\frac{\Omega r \sin\theta_{\rm pk}}{c}\right]^2}
\ee
We show $\Gamma_{\rm d}(r)-1$ as a dotted black curve in the right panel of Fig.~\ref{fig:coronal_region_bulk_components}. This matches fairly well with our measurement of the mean bulk Lorentz factor weighted by the radial energy flux (yellow dashed curve in the right panel of Fig.~\ref{fig:coronal_region_bulk_components}).   

So far we have characterized the mean bulk motions. Within the paradigm of cold-chain Comptonization, the properties of the Comptonized hard nonthermal X-ray emission are determined by the stochasticity in the plasmoid motions. To this end, we compute the polar- and time-averaged bulk energy spectra within the dissipative jet sheath, in the frame moving with the local mean velocity (as quantified above). The histogram of bulk motions measured in this frame is shown by the solid curves in the left panel of Fig.~\ref{fig:coronal_region_bulk_motions} as a function of radial distance, from $5\,R_{\rm g}$ (blue) to $40\,R_{\rm g}$ (red) in increments of $5\,R_{\rm g}$. The distribution of stochastic bulk motions resembles a Maxwellian with an `effective bulk temperature' of $\sim200$\,keV at $5\,R_{\rm g}$,  decreasing down to $\sim100$\,keV at $40\,R_{\rm g}$ (see dashed and dotted black curves). In the right panel, we show the average bulk Lorentz factor $\langle\Gamma^\prime-1\rangle$ and its standard deviation $\sigma_{\Gamma^\prime-1}$ (as shown by the dark grey band), both measured in the frame moving with the local mean velocity. We find that $\langle\Gamma^\prime-1\rangle\pm\sigma_{\Gamma^\prime-1}$ decreases from $0.40\pm0.14$ at $r=10\,R_{\rm g}$ to $0.3\pm0.1$ at $r=40\,R_{\rm g}$.

In the traditional picture of thermal Comptonization, hot electrons with a temperature of $\sim100$\,keV are required in order to explain the observed hard nonthermal X-ray spectra. In our model, it is the stochasticity in the bulk motions of dissipative structures within the jet sheath---remarkably, having an effective bulk temperature of $\sim100$\,keV---that allows soft photons to be Compton scattered to a hard nonthermal X-ray distribution. In previous works, we have employed  Monte Carlo radiative transfer calculations to show that bulk motions with an effective bulk temperature of $\sim100$\,keV are capable of explaining the observed hard nonthermal X-ray spectra through the cold-chain Comptonization process (\citetalias{Sridhar+21c}, \citetalias{Sridhar+23a}). In local simulations of relativistic reconnection, the stochasticity of bulk motions is a natural by-product of the plasmoid chain dynamics (e.g., \citetalias{Sridhar+21c}, \citetalias{Sridhar+23a}). Here, our choice of $\sigma_{\rm h}>0.15$ and $\beta>0.1$ to identify dissipative regions in the jet sheath is tailored toward capturing reconnection layers and plasmoids. Still, stochastic motions may be contributed by other sources, e.g., the waves/vortices seen at the jet sheath (check panels [a,b] of Fig.~\ref{fig:theta-time_parameters} for the presence of nonlinear waves along the jet sheath).

\subsection{Optical depth} \label{subsec:tau}

The results we have presented so far from our GRRMHD simulations are scale-free. To obtain observable properties, we consider parameters typical of Cygnus X-1 in the hard state: a black hole of mass $M_\bullet=20M_\odot$ with an X-ray luminosity of $2\times10^{37}\,{\rm erg\,s^{-1}}$---corresponding to an accretion power of $\dot{M} c^2 \sim 10^{38}\,{\rm erg\,s^{-1}}$ assuming an efficiency of $\xi=0.2$ (see also \S\ref{subsec:Poynting_flux} and Appendix~\ref{Appendix:dissipation}). Taking the simulation time unit to be $T_{\rm sim}=GM_\bullet/c^3$, the length unit to be $L_{\rm sim}=GM_\bullet/c^2$, and the average accretion rate (in dimensionless simulation units) to be $\dot{M}_{\rm sim}\sim 10$ (see Fig.~\ref{fig:horizon_params}), we obtain the simulation mass unit of $M_{\rm sim}=\dot{M} T_{\rm sim}/\dot{M}_{\rm sim}\simeq1.1\times10^{12}$\,g, and the magnetic field strength unit of $B_{\rm sim}=c\sqrt{4\pi M_{\rm sim}/L_{\rm sim}^3}\simeq2.2\times10^7$\,G.\footnote{Here, we explicitly include the values of $4\pi$ and $c$ to calculate the observationally relevant quantities in CGS units.} 

Any coronal model must satisfy the requirement that the Thomson optical depth $\tau_{\rm T}\sim 0.1-1$ in order to explain the shape of the nonthermal X-ray spectrum seen during the hard state of X-ray binaries \citep{Zdziarski+98, Garcia+15, Sridhar+20}. We use the scaling parameters derived above to compute the expected optical depth across the dissipative jet sheath of our GRRMHD simulation. We note, however, that our estimate of optical depth is rather rudimentary since our simulation includes neither self-consistent radiative physics (e.g., cooling, pair production) nor proper electron thermodynamics (e.g., electron-ion Coulomb coupling). Future dedicated simulations, following \citet{Liska+22}, will be needed to provide a more detailed assessment of the composition (pair plasma vs electron-proton plasma) and the optical depth of the dissipative jet sheath. 

Given that our simulation cannot inform us of the sheath composition, we will treat separately the case of electron-proton plasma and electron-positron plasma. If the sheath is efficiently loaded with baryons from the accretion disk, the optical depth across the jet sheath (i.e., in the polar direction) can be expressed as \citep{belo_17}
\be \label{eq:tau_ep}
\tau_{\rm e^-p} \sim \left(\frac{w}{0.1\,r}\right)\,\frac{2\ell_{\rm B}/\sigma_{\rm c}}{m_{\rm p}/m_{\rm e}},
\ee
where $\ell_{\rm B}$ is the magnetic compactness defined as,
\be \label{eq:compactness}
\ell_{\rm B} = \frac{\sigma_{\rm T}}{m_{\rm e}c^2} \frac{B_{\rm sim}^2}{8\pi} \langle b^2\rangle r,
\ee
for a comoving field $b$ in simulation units.
In Eq.~(\ref{eq:tau_ep}), we assume the width of the current sheets in the jet sheath to be $w\sim0.1\,r$. In the case of plasmoid chain Comptonization, \citet{belo_17} adopted a similar definition, where $0.1$ was the aspect ratio of the reconnection layer, set by the reconnection rate.  We remark that Eq.~(\ref{eq:tau_ep}) is a reasonable estimate of optical depth only if the mass density in the sheath  (in $\sigma_{\rm c}$) is not dominated by the density floors of the GRRMHD algorithm. 

\begin{figure}
    \includegraphics[width=\linewidth]{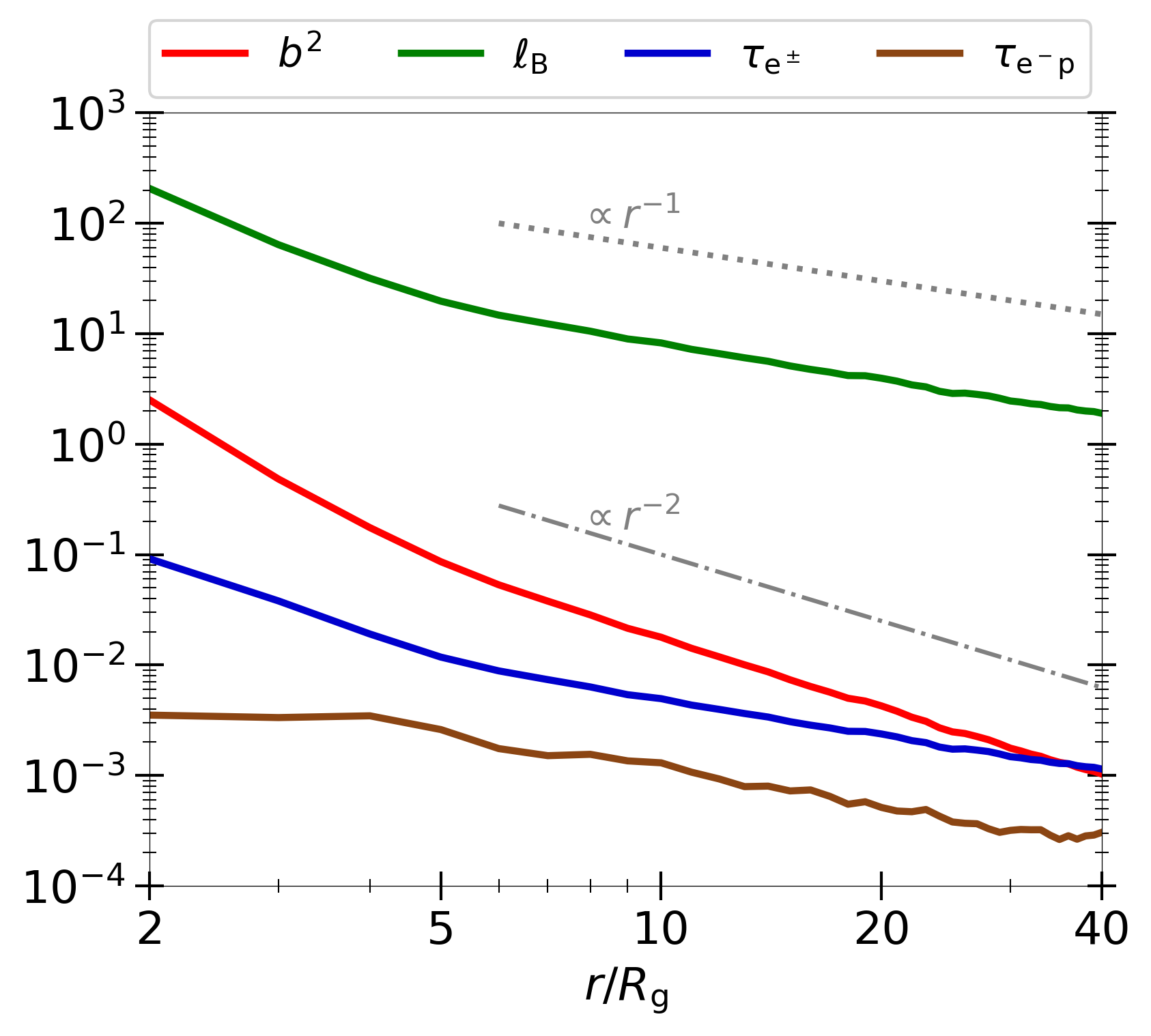}
    \caption{Polar- and time-averaged radial profiles of the comoving magnetic energy density (red), the magnetic compactness (green), the optical depth of electron-positron pairs (blue) and of electron-proton plasma (brown) in the dissipative jet sheath, as obtained from our simulation scaled to the
    parameters typical of Cygnus X-1 (see \S\ref{subsec:tau} for more details). 
    }
    \label{fig:tau}
\end{figure}

In X-ray binaries such as Cygnus X-1, $\sim1\%$ of the X-ray luminosity is observed in the MeV band \citep{McConnell+02}, which may give rise to $e^\pm$ pair creation through photon-photon annihilation. Following the radiative PIC studies of reconnection (\citetalias{Sironi_20}, \citetalias{Sridhar+21c}, \citetalias{Sridhar+23a}), we assume that a fraction $f_{\rm HE}\sim 0.3$ of the magnetic energy is converted to high-energy particles. If synchrotron losses are negligible, a fraction $f_{\pm}\sim0.1$ of these particles' energy converts to the rest-mass energy of secondary $e^\pm$ pairs \citep{Svensson_87}. When $u\equiv (3/16)f_{\pm}f_{\rm HE}\ell_{\rm B}<1$, pair creation is balanced by escape. On the other hand, the annihilation balance is approached when $u>1$. The magnetic compactness measured in the dissipative jet sheath (green curve in Fig.~\ref{fig:tau}) is such that $u>1$ for $r\lesssim 3 \, R_{\rm g}$, and $u<1$ for $r\gtrsim 3 \, R_{\rm g}$. In this case, the optical depth of a self-regulated pair plasma can be estimated to be \citep{belo_17},
\begin{align}
\tau_{\rm e^\pm} \sim\frac{16}{3}\left(\frac{\xi}{0.2}\right)\left(\frac{w}{0.1\,r}\right)\times
\Bigg\{\begin{array}{ll}
u,        & u < 1 \\
\sqrt{u}, & u > 1
\end{array}
,
\label{eq:tau_ee}
\end{align}
where we adopt $\xi=0.2$ as the efficiency of electromagnetic dissipation (see \S\ref{subsec:Poynting_flux} and Appendix~\ref{Appendix:dissipation}). Fig.~\ref{fig:tau} shows the polar-averaged and time-averaged profiles of $\ell_{\rm B}$, $\tau_{\rm e^-p}$ and $\tau_{\rm e^\pm}$ as a function of radial distance from the black hole, considering only regions belonging to the dissipative jet sheath (i.e. where $\beta>0.1$ and $\sigma_{\rm h}>0.15$). We find that, for a (putative) corona at $2\lesssim r/R_{\rm g}\lesssim 20$, the electron-proton optical depth is $\tau_{\rm e^-p}\sim 10^{-3}-10^{-2}$, while the self-regulated pair optical depth is $\tau_{\rm e^\pm}\sim 0.01-0.1$. Our estimate of $\tau_{\rm e^-p}$ is affected by the threshold $\sigma_{\rm h}>0.15$ used to identify the dissipative jet sheath; a less restrictive threshold will likely give higher $\tau_{\rm e^-p}$.
We note that future global GRRMHD simulations with self-consistent radiative physics are needed for a proper estimate of the optical depth in the dissipative jet sheath.

\section{Discussion} \label{sec:summary}

Nonthermal X-ray emission is seen from active galactic nuclei \citep[AGNs; ][]{Fabian+15} and X-ray binaries \citep{Done+07}. In previous papers (\citetalias{Sironi_20}, \citetalias{Sridhar+21c}, \citetalias{Sridhar+23a}, \citealt{Gupta+24}), we performed kinetic plasma (PIC) simulations of relativistic and trans-relativistic magnetic reconnection in $e^\pm$ and electron-ion plasmas. The simulations demonstrate that the popular scenario of hard X-ray production by thermal Comptonization is unfeasible for a corona heated by magnetic reconnection, since most of the electrons in the reconnection layer are kept at low temperatures by the strong Compton cooling. However, the reconnection layer can still produce the observed hard X-ray spectra via Comptonization by the stochastic bulk motions of the Compton-cooled plasmoid chain \citep{belo_17}. In this paper, we employ global GRRMHD simulations to explore which regions can produce the stochastic trans-relativistic plasmoid chain that could act as a Comptonizing corona. Our main results are as follows.

\begin{enumerate}

\item The sheath between the jet and the surrounding medium is an important site of energy dissipation, as revealed by the copious presence of reconnection layers and plasmoid chains. 
This region (which we have called the `dissipative jet sheath') is well identified by a combination of thresholds in plasma beta, $\beta>0.1$, and hot magnetization, $\sigma_{\rm h}>0.15$, i.e., more magnetized than the accretion disk yet less magnetized than the jet. 

\item The field lines that reconnect in this region are primarily radial. The time-averaged azimuthal field $B_{\rm out}$ in the dissipative jet sheath is comparable in strength to the reconnecting fields. While $B_{\rm out}$ does not always act as a guide field, we find that its polarity often stays constant across the current sheets located in the jet sheath (i.e., it can act as a guide field).

\item The radial electromagnetic power flowing in the jet sheath is twice as large as the electromagnetic power flowing in the jet core and is comparable to the overall accretion power. The electromagnetic power in the jet sheath decreases by $\sim 20\%$ from 2 to $10\,R_{\rm g}$. This may be able to satisfy the energetic requirements to explain the nonthermal X-rays from the hard state of black hole X-ray binaries \citep{McClintock&Remillard_06, Done+07}. We find that the polar angle of the jet sheath depends on distance as $\theta_{\rm pk}\simeq0.3\pi(r/5\,R_{\rm g})^{-0.5}$, indicating a paraboloidal surface. Note that the X-ray-radio flux correlation ($F_{\rm radio}\propto F_{\rm X-ray}^{0.6}$) seen during the hard state, and the observed time lags were recently explained by invoking an outflowing, parabolic Comptonizing region \citep{Kylafis+23, Kylafis&Reig_2024}.

\item The mean bulk motions in the jet sheath are trans-relativistic and dominated by the radial component, with time-averaged dimensionless 4-velocity $\langle \Gamma\beta_{\rm r} \rangle=1.2$. In the frame moving with the local mean velocity, the distribution of stochastic bulk motions resembles a Maxwellian with an effective bulk temperature of ${\cal O}$(100)\,keV, which is required to produce a Comptonized nonthermal X-ray spectrum in agreement with the observations. While the presence of trans-relativistic plasmoids from magnetic reconnection is a sufficient condition for stochastic bulk motion Comptonization, we emphasize that the stochasticity can also be potentially brought about by other means such as trans-relativistic turbulence \citep[e.g.,][]{Groselj+24} and/or Kelvin-Helmholtz-like vortices that are expected along the jet sheath \citep[e.g.,][]{Wong+21,chow_23,Davelaar+23}.

\item The trans-relativistic radial bulk velocity of the jet sheath should have implications for the strength of the disk-reflected component \citep{Beloborodov_99}, the polarization angle (PA), and the PD. If the corona is launched away from the disk with speed $0.1\lesssim v/c\lesssim0.8$---our measurements are in this range---\citet{Beloborodov_98} found that the PA would be parallel to the disk normal---this is the PA measured by IXPE for Cygnus X-1 \citep{Krawczynski+22}. Furthermore, IXPE observations revealed a large PD ($\sim4\%$), which could be explained by an outflowing Comptonizing region with speed $v/c\gtrsim0.4$ \citep[akin to model (c) of ][and in the range of our measurements]{Poutanen+23}. \cite{Dexter&Begelman24} showed that the observed PA and PD of Cygnus X-1 could be explained by invoking a bulk outflow with $\Gamma\gtrsim1.5$ along a conical sheath. Their requirement that the cone half-opening angle be $0.15\pi\lesssim\theta_{\rm cone}\lesssim0.3\pi$ means that, in our model, most of the Comptonization must occur between 5 and 20\,$R_{\rm g}$ (Fig.~\ref{fig:ExBr_theta_time_f_csjet}[c]).

\item We estimate the Thomson optical depth across the dissipative jet sheath to be $\tau\sim0.01-0.1$ at $2\lesssim r/R_{\rm g}\lesssim 10$. A proper assessment of the optical depth and of the composition of the dissipative jet sheath (electron-proton vs. electron-positron plasma) requires dedicated radiative GRRMHD simulations, which are beyond the scope of this work. 
\end{enumerate}

Several lines of observational evidence have pointed toward a link between the presence of a compact radio jet and the emission of nonthermal X-rays as seen during the low/hard states of X-ray binaries and AGN \citep{Fender&Kuulkers_01, Gallo+03, Fender+04b, Gallo+18, Mendez+22}. Our candidate corona---the dissipative jet sheath---is naturally linked to the presence of a jet. As already discussed, our work is motivated by the radiative GRMHD simulations of \citet{Dexter+21, Liska+22}, which showed the formation of an inner thicker disk (within $\sim10\,R_{\rm g}$) with a strong poloidal field (akin to a MAD state) and a powerful jet, in case the flow is initially threaded with a dominant poloidal flux at large radii. 
It has been argued that the accretion disk might not be MAD during bright hard states based on the presence of Type-C quasi-periodic oscillations, which are thought to be associated to the precession of a thin disk \citep{Ingram&Motta_19}, and so not expected from a thicker, MAD disk \citep[][]{Fragile+23}. We note, however, that Cygnus X-1 does not exhibit strong Type-C quasi-periodic oscillations during its hard state. 

We conclude by emphasizing limitations of the analysis presented in this paper. Our results are based on two-dimensional MHD simulations of accretion onto a highly spinning black hole in the MAD regime, i.e., with strong poloidal flux near the disk's inner edge. We expect that the proposed Comptonization mechanism will also work in three-dimensional (3D) simulations, where both the kinetic drift-kink instability and the MHD flux-rope kink instability are enabled. This expectation is supported by existing local-box radiative 3D PIC simulations \citep{Sironi_20}. In the future, it may be useful to investigate how our conclusions are influenced by: (1) three-dimensionality in global simulations; (2) a lower and possibly retrograde spin of the black hole; (3) a dominant toroidal component of the magnetic field in the disk, which may weaken the jet; (4) a spatially-dependent prescription for resistivity based on kinetic simulations of collisionless reconnection \citep{Selvi+23, Bugli+24}, which might yield a faster reconnection rate and may affect the properties of the plasmoid bulk motions in the jet sheath; and (5) the self-consistent interaction between plasma and radiation. The latter is essential to properly determine the plasma composition, the resulting optical depth, and self-consistent Comptonized X-ray spectra.

\section*{Acknowledgements}

This work benefits from the discussions with Roger Blandford, Andy Fabian, Kyle Parfrey, Oliver Porth, Jim Stone, Sasha Tchekhovskoy, Chris Thompson, and members of the Simons Collaboration on Extreme Electrodynamics of Compact Sources (SCEECS).

N.S. acknowledges the support from NASA (grant No. 80NSSC22K0332), NASA FINESST (grant No. 80NSSC22K1597), Columbia University Dean’s fellowship, and a grant from the Simons Foundation. B.R. is supported by the Natural Sciences \& Engineering Research Council of Canada (NSERC). L.S. acknowledges support from the DoE Early Career Award DE-SC0023015. This work was supported by a grant from the Simons Foundation (MP-SCMPS-00001470) to L.S. and B.R., and facilitated by Multimessenger Plasma Physics Center (MPPC), grant PHY-2206609 to L.S. J.D. is supported by NASA through the NASA Hubble Fellowship grant HST-HF2-51552.001A, awarded by the Space Telescope Science Institute, which is operated by the Association of Universities for Research in Astronomy, Incorporated, under NASA contract NAS5-26555. A.M.B. acknowledges support by NASA grants 80NSSC24K1229 and 21-ATP21-0056, NSF grant AST-2408199, and Simons Foundation grant No. 446228.

Part of this research was carried out at, and supported by, the Munich Institute for Astro-, Particle and BioPhysics (MIAPbP), which is funded by the Deutsche Forschungsgemeinschaft (DFG, German Research Foundation) under Germany's Excellence Strategy – EXC-2094 – 390783311. N.S. performed part of this work at the Aspen Center for Physics, which is supported by the National Science Foundation grant PHY-2210452. 

The computational resources and services used in this work were partially provided by Columbia University (Habanero, Terremoto, and Ginsburg HPC clusters), facilities supported by the Scientific Computing Core at the Flatiron Institute, a division of the Simons Foundation, and by the VSC (Flemish Supercomputer Center), funded by the Research Foundation Flanders (FWO) and the Flemish Government – department EWI.

\newpage
\bibliographystyle{aasjournal}
\bibliography{references}

\appendix

\begin{figure*}
    \centering
    \includegraphics[width=\textwidth]{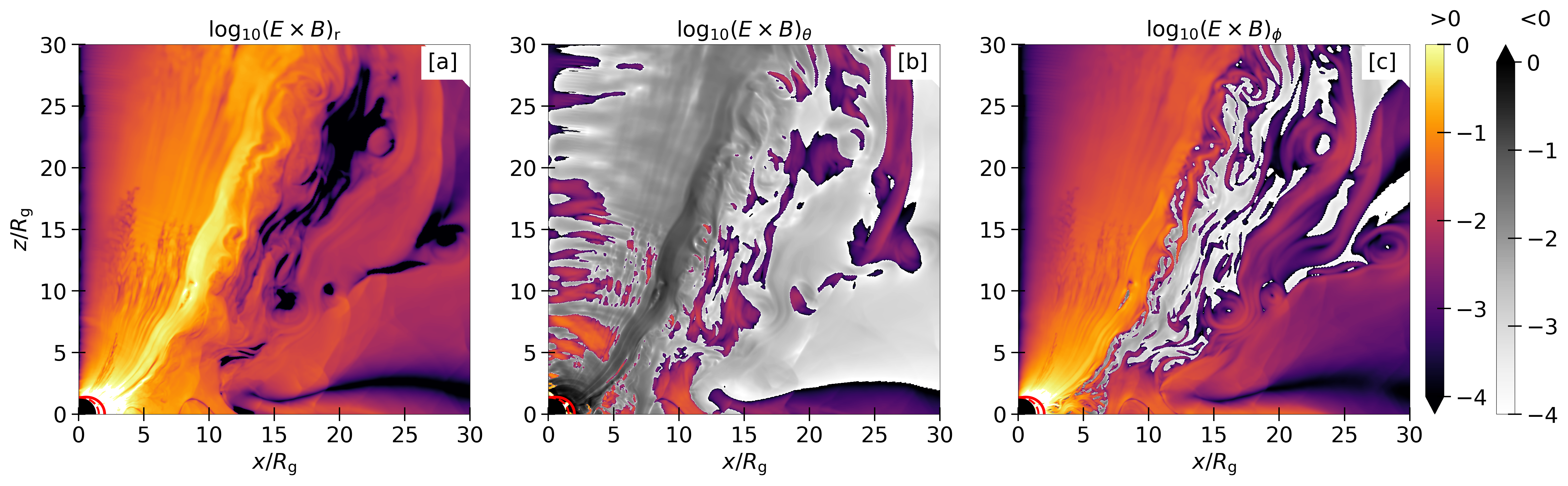}
    \caption{Panels [a,b,c] show the radial, polar, and toroidal components of the Poynting flux at $Tc/R_{\rm g}=2870$. In all panels, colors denote values of the Poynting flux $>0$, while the greyscale denotes values $<0$.}
    \label{fig:ExB_components}
\end{figure*}

\section{Components of the Poynting flux} \label{Appendix:Poynting_components}
\begin{figure*}
    \centering
    \includegraphics[width=\textwidth]{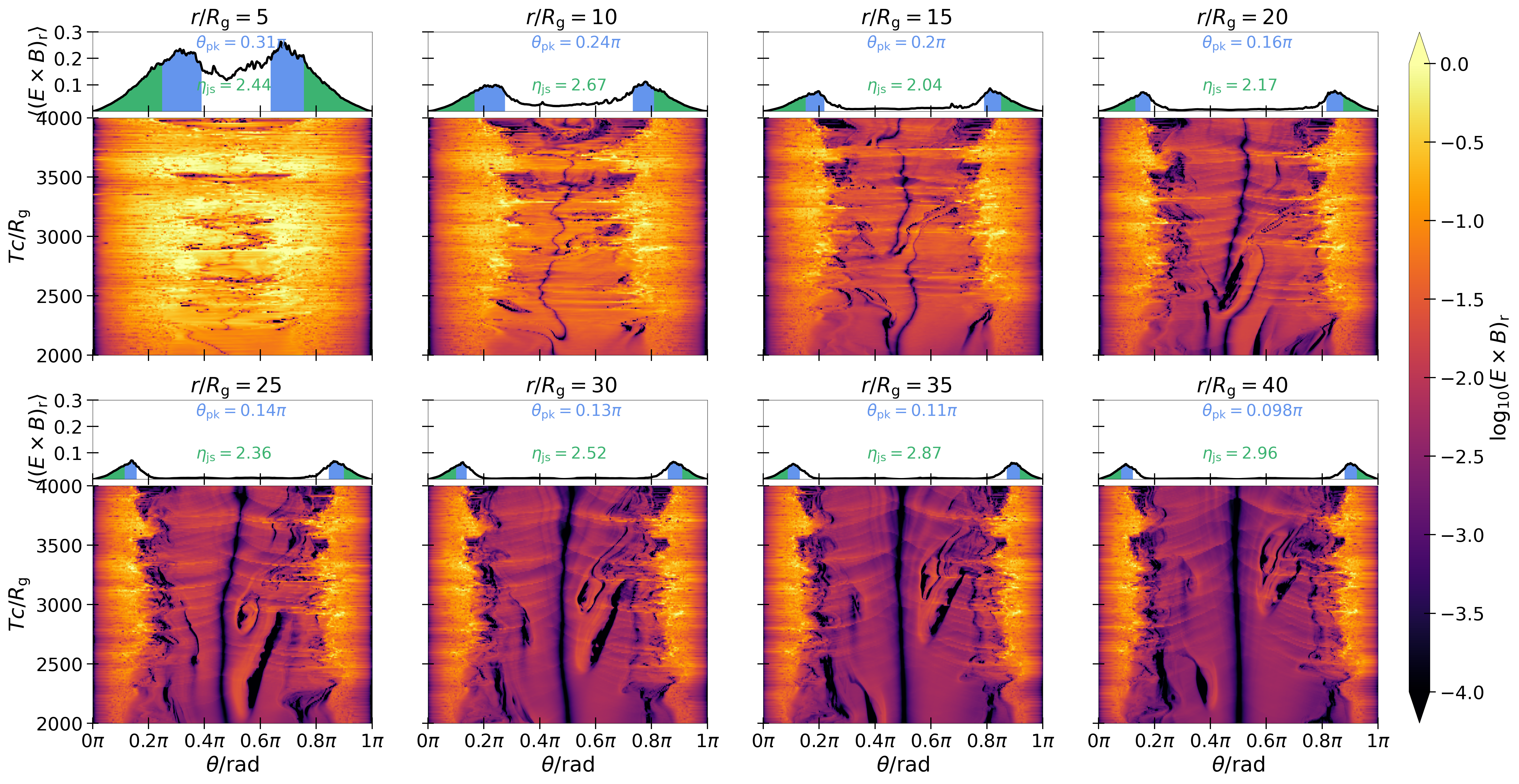}
    \caption{Time-evolution of the radial Poynting flux as a function of $\theta$. Each panel denotes a different radius. All the quantities shown are the same as in Fig.~\ref{fig:ExBr_theta_time_f_csjet}[b].}
    \label{fig:ExBr_theta_time_radii}
\end{figure*}

In Fig.~\ref{fig:ExB_components}, we show the covariant radial, polar, and azimuthal components of the Eulerian Poynting flux in panels [a, b, c], respectively, at time $Tc/R_{\rm g}=2870$. In the jet sheath, we find the radial component of the Poynting flux to be dominant. In Fig.~\ref{fig:ExBr_theta_time_radii}, we show the time-evolution of $(E\times B)_{\rm r}$ as a function of the polar angle ($\theta$) for different radii in the bottom sub-panels. All the quantities are the same as in Fig.~\ref{fig:ExBr_theta_time_f_csjet}[b]. The radial Poynting flux is seen to decrease as the distance increases (mostly due to the larger surface area); a deeper analysis of this is presented below and in the main text (\S\ref{subsec:Poynting_flux}).

\section{Jet-sheath efficiency} \label{Appendix:Poynting_efficiency}
\begin{figure*}
    \centering
    \includegraphics[width=\textwidth]{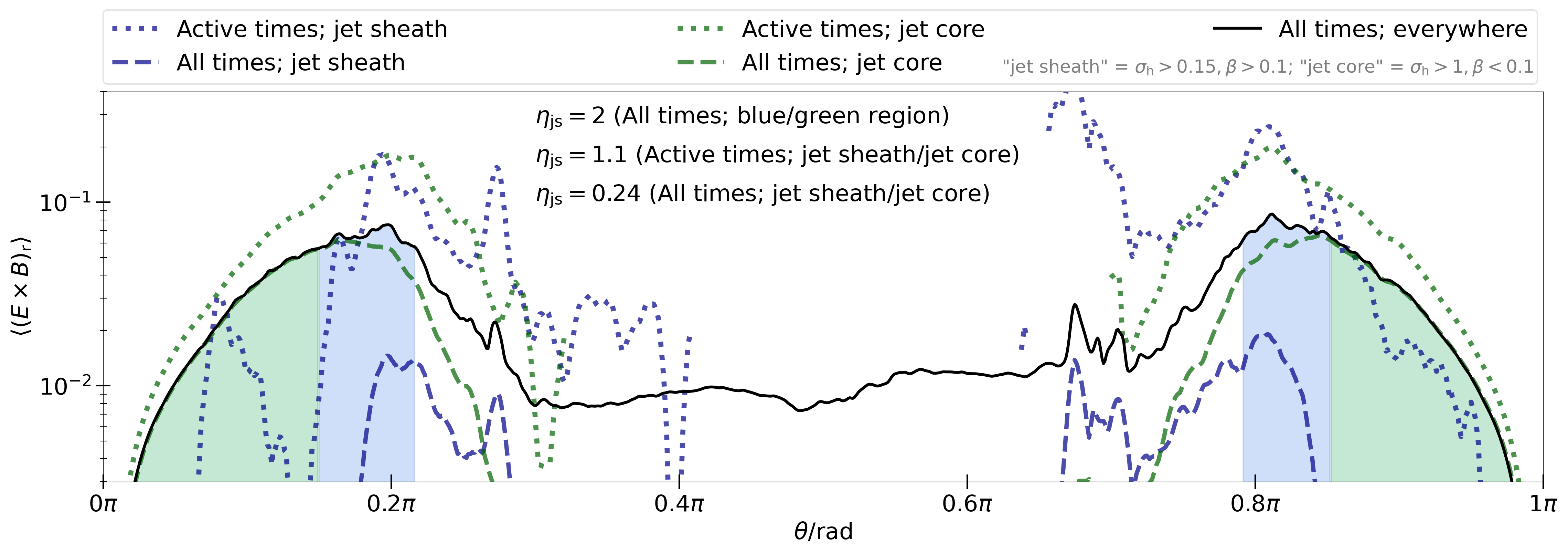}
    \caption{Time-averaged radial Poynting flux $\langle (E\times B)_{\rm r} \rangle$ as a function of $\theta$ calculated at the fiducial radius of $r=15\,R_{\rm g}$. The blue and green dotted curves denote $\langle (E\times B)_{\rm r} \rangle$ computed only at the times when the criteria $\sigma_{\rm h}>0.15, \beta>0.1$ for jet sheath, and $\sigma_{\rm h}>1, \beta<0.1$ criterion for jet core are attained. The blue and green dashed curves denote $\langle (E\times B)_{\rm r} \rangle$ averaged over all times during the quasi-steady state. For dotted and dashed green lines, we consider only regions that satisfy $\sigma_{\rm h}>1, \beta<0.1$ (jet core). For dotted and dashed blue lines, we consider only regions that satisfy $\sigma_{\rm h}>0.15, \beta>0.1$ (jet sheath). The black curve denotes $\langle (E\times B)_{\rm r} \rangle$ averaged over all times in the whole volume; the blue and green patches define the jet sheath and jet core regions as done in the main text. The jet sheath efficiency $\eta_{\rm js}$ obtained for different definitions of jet sheath and jet core is also listed (see Appendix~\ref{Appendix:Poynting_efficiency} for more details).}
    \label{fig:ExBr_f_csjet_options}
\end{figure*}

In Fig.~\ref{fig:ExBr_f_csjet_options}, we check how the jet sheath efficiency $\eta_{\rm js}$ depends on different definitions for jet sheath and jet core, at our fiducial radius of $r=15\,R_{\rm g}$. The efficiency $\eta_{\rm js}$ presented in the main paper is defined as the ratio of polar-angle-integrated $\langle (E\times B)_{\rm r} \rangle$ in the jet sheath region to the jet core region. The solid black curve in Fig.~\ref{fig:ExBr_f_csjet_options} shows $(E\times B)_{\rm r}$ averaged during the quasi-steady state, as a function of $\theta$. One way we define the jet sheath is such that $\langle (E\times B)_{\rm r} \rangle[\theta] \geq 0.75\langle (E\times B)_{\rm r} \rangle[\theta_{\rm pk}]$, where $\theta_{\rm pk}$ is the polar angle where $\langle (E\times B)_{\rm r}\rangle $ peaks in the upper and lower hemispheres (blue patches); the jet core is the remaining green patch. For example, the ranges of $\theta$ for the jet sheath region in the upper hemisphere going by this definition are: 
$0.25\pi\lesssim \theta\lesssim0.39\pi$ at 5\,$R_{\rm g}$, 
$0.17\pi\lesssim \theta\lesssim0.27\pi$ at 10\,$R_{\rm g}$, 
$0.15\pi\lesssim \theta\lesssim0.22\pi$ at 15\,$R_{\rm g}$, 
$0.13\pi\lesssim \theta\lesssim0.18\pi$ at 20\,$R_{\rm g}$, 
$0.11\pi\lesssim \theta\lesssim0.16\pi$ at 25\,$R_{\rm g}$, 
$0.10\pi\lesssim \theta\lesssim0.14\pi$ at 30\,$R_{\rm g}$,
$0.09\pi\lesssim \theta\lesssim0.13\pi$ at 35\,$R_{\rm g}$, 
and $0.08\pi\lesssim \theta\lesssim0.12\pi$ at 40\,$R_{\rm g}$. This definition yields $\eta_{\rm js}\sim2$, as discussed in the main paper. For other purposes, in the paper, we have defined the dissipative jet sheath as the region having $\sigma_{\rm h}>0.15, \beta>0.1$; regions satisfying these conditions only exist at certain times (which we shall call `active times'). We complement this by defining here the jet core as the region with $\sigma_{\rm h}>1, \beta<0.1$. The $\langle (E\times B)_{\rm r} \rangle$ averaged only during active times is shown by blue (jet sheath) and green (jet core) dotted curves. The corresponding jet sheath efficiency is $\eta_{\rm js}\sim1.1$ (here, definitions of the jet sheath and jet core are based on our $\sigma_{\rm h}$ and $ \beta$ thresholds). This would be $\eta_{\rm js}\sim0.24$ if averaged over `all times' during the quasi-steady state (dashed curves). We also find that the jet sheath efficiency increases by a factor of $\sim2$ by considering the total radially flowing power $ (E\times B)_{\rm r} (1+1/\sigma_{\rm h}) $ instead of just the electromagnetic component $(E\times B)_{\rm r}$. 

\section{Electromagnetic dissipation along the jet sheath} \label{Appendix:dissipation}
\begin{figure*}
    \centering
    \includegraphics[width=0.6\textwidth]{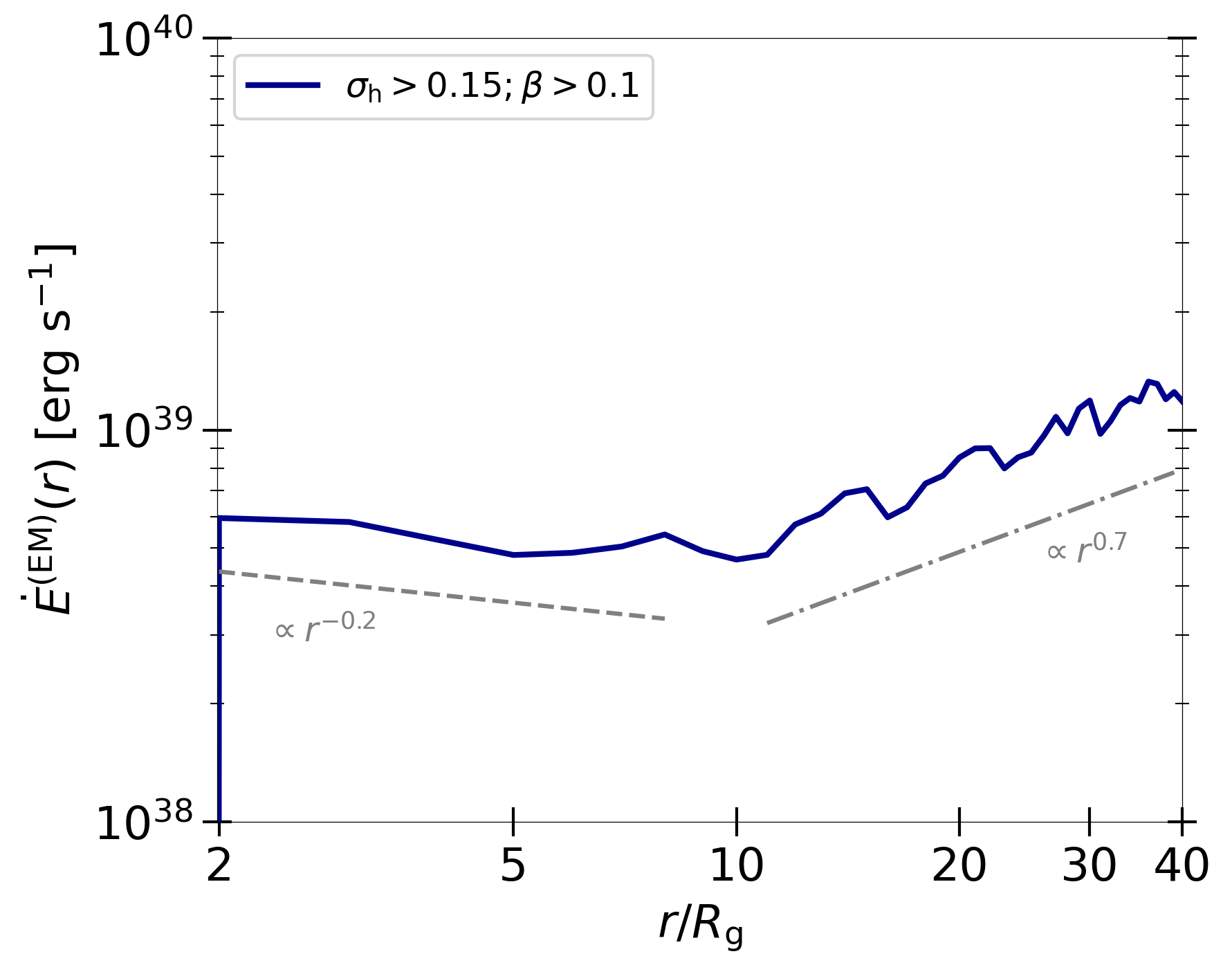}
    \caption{Time-averaged radially flowing electromagnetic power, $\dot{E}^{\rm (EM)}(r) = 2\pi\int\sqrt{g^{\rm rr}}\langle E\times B\rangle_{\rm r}\sqrt{-g}{\rm d}\theta$ (for Cygnus X-1 parameters) as a function of radius, along the jet sheath defined by our $\sigma_{\rm h}>0.15$ and $\beta>0.1$ thresholds.}
    \label{fig:ExBr_radial_corona}
\end{figure*}

Furthermore, we estimate the fraction of the radially-flowing electromagnetic power that is dissipated in the jet sheath: we find that the radially-flowing electromagnetic power in the jet sheath $\dot{E}^{\rm (EM)}(r) = 2\pi\int\sqrt{g^{\rm rr}}(E\times B)_{\rm r}\sqrt{-g}{\rm d}\theta$ decreases with distance as $ \propto r^{-0.2}$ between 2 and 10\,$R_{\rm g}$ i.e., it decreases by $\sim 20\%$ in our candidate coronal region from $2-10\,R_{\rm g}$. We associate this fraction $\xi=0.2$ with the efficiency of converting the electromagnetic energy to plasma (and radiation), and find up to $\sim10^{38}\,{\rm erg\,s^{-1}}$ to be dissipated between 2 and 10\,$R_{\rm g}$ (for the parameters typical of Cygnus X-1; see \S\ref{subsec:tau}). This is shown in Fig.~\ref{fig:ExBr_radial_corona}; the dark blue solid curve is computed from the regions of the jet sheath defined by $\sigma_{\rm h}>0.15$ and $\beta>0.1$ (time-averaged during the active times, as mentioned above). As discussed by \cite{Qian+18}, the increasing $\dot{E}^{\rm (EM)}(r)$ at larger radii $r\gtrsim10\,R_{\rm g}$ could be due to the contribution of large-angled disk winds. This trend is robust regardless of whether the dissipative jet sheath is defined based on the $\sigma>0.15, \beta>0.1$ threshold or the $\langle (E\times B)_{\rm r} \rangle[\theta] \geq 0.75\langle (E\times B)_{\rm r} \rangle[\theta_{\rm pk}]$ condition.

\end{document}